%% file: main.tex
\documentclass[twocolumn]{emulateapj}
\usepackage{amssymb,graphicx}
\usepackage{epsfig}
\usepackage{psfrag}
\usepackage[usenames]{color}
\usepackage{multirow}
\usepackage{rotating}

\usepackage{graphicx,color}
\usepackage{amsmath,amssymb}
\usepackage{verbatim}
\usepackage{float}
\usepackage{wasysym}
\usepackage{dsfont}
\usepackage{amsfonts}
\usepackage{mathrsfs}
\usepackage{multirow}
\usepackage{times}
\usepackage{bm}
\usepackage{hyperref}
\usepackage{scalefnt}
\usepackage{xspace}
\usepackage{soul}
\usepackage{lipsum} 
\usepackage{physics}
\hypersetup{
  colorlinks=true,        
  linkcolor=blue,         
  citecolor=cyan,         
}
\usepackage{blindtext, rotating}
\usepackage{pifont}
\usepackage[normalem]{ulem}   
\usepackage{grffile}
\usepackage{booktabs}
\graphicspath{{./}}        

\input kigou.tex

\newcommand{\be}{\begin{equation}}
\newcommand{\ee}{\end{equation}}

\def\QEQ{{%
    \setbox0\hbox{$I$}%
    \rlap{\hbox to \wd0{\hss--\hss}}\box0
}}

\shorttitle{Influence of rotation on magnetic field stability in NS}
\shortauthors{Venturi, Yip, Cheong \& Ruiz}

\begin{document}
%
\title{Impact of rotation on magnetic field stability and orientation in isolated neutron stars}
%
%
%
\author{Fabrizio Venturi Piñas$^1$, Anson Ka Long Yip$^2$, Patrick Chi-Kit Cheong$^{3,4,5}$, Milton Ruiz$^{1}$}
\affil{
$^1$Departamento de Astronom\'{\i}a y Astrof\'{\i}sica, Universitat de Val\'encia, Dr. Moliner 50, 46100, Burjassot (Val\`encia), Spain\\
$^2$Department of Physics, The Chinese University of Hong Kong, Shatin, N.T., Hong Kong\\
$^3$Department of Physics, University of California, Berkeley, Berkeley, CA 94720, USA\\
$^4$Center for Nonlinear Studies, Los Alamos National Laboratory, Los Alamos, NM 87545, USA\\
$^5$Department of Physics \& Astronomy, University of New Hampshire, 9 Library Way, Durham NH 03824, USA
}

\begin{abstract}
Neutron stars are the most compact horizonless objects in the Universe, exhibiting the strongest known magnetic fields.
They are potential sources of coincident gravitational waves and electromagnetic radiation across the entire spectrum.
However, the internal configuration of their magnetic fields and the mechanisms that stabilize them remain open questions.
As a step forward in understanding the timescale for the emergence of magnetic instabilities that disrupt stellar field
configurations, we study the impact of stellar rotation using three-dimensional general relativistic numerical simulations of 
uniformly rotating, isolated neutron stars threaded by strong, poloidal, pulsar-like magnetic fields. The initial stellar configurations 
assume perfect conductivity and are stationary and axisymmetric. We explore a range of angular velocities, from non-rotating 
stars to those near the mass-shedding limit. We find that the stars spontaneously develop differential rotation, which triggers 
the appearance of a strong toroidal magnetic field component. Non-rotating neutron stars are unstable to the Tayler and Parker 
instabilities, which significantly change the magnetic field geometry. These instabilities lead to a rapid reduction of 
the initial magnetic energy by $\sim 99\%$ within $\sim 4$ Alfv\'en times of their onset. In contrast, rotation 
significantly delays the development of these instabilities and, in some cases, mitigates their effects. Highly rotating
models retain up to $\sim 30\%$ of their magnetic energy for at least $\sim 10$ Alfv\'en times. Our results suggest that rotation 
plays a crucial role in stabilizing the magnetic field of neutron stars, regardless of its initial configuration.
\end{abstract}

\maketitle

\section{Introduction}
\label{Sec:Intro}
Neutron stars (NSs)
are formed from the remnants of massive stars that have undergone supernova explosions~\citep{2002RvMP...74.1015W,Lattimer:2000nx,1995ApJ...445L.129B}.
These stars exhibit extraordinarily strong magnetic fields and can be classified into several categories: 
i) old-recycled pulsars with surface magnetic fields around $10^8\,\rm G$~\citep{1994ApJ...421L..15C,1997MNRAS.284..311K}; 
ii) typical pulsars with fields around $10^{12}\,\rm G$~\citep{1997MNRAS.284..311K,Bhattacharya1991}; and 
iii) magnetars, with fields reaching up to or exceeding $10^{15}\,\rm G$~\citep{1992ApJ...392L...9D,1993ApJ...408..194T,1996ApJ...473..322T}.
There are at least two leading hypotheses to explain the formation of such powerful magnetic fields: 
i) the fossil field origin, where the magnetic field is inherited from the progenitor star~\citep{Ferrario:2006ib,Spruit:2007bt,Braithwaite:2004ubj};
and  ii) a dynamo mechanism operating during the proto-neutron star stage~\citep{1996ApJ...473..322T,Reisenegger:2003pj,2005PhR...417....1B}.
While modest growth of the magnetic field can also occur due to thermoelectric effects~\citep{Vigano:2021olr,1996ApJ...473..322T}, these processes
are unlikely to generate the extreme magnetar-scale fields.

These strong magnetic fields play a key role in the observable phenomena of NSs, such as the magnetic dipole radiation seen in pulsars and
the magnetically-powered burst activity of magnetars~\citep{1969ApJ...157..869G,1995MNRAS.275..255T}.
They are typically invoked to explain the quasi-periodic oscillations detected following magnetar giant flares (see e.g.
\cite{Gabler:2010rp,Sotani:2006at}). Additionally, magnetic fields may induce deformations that could lead to significant gravitational wave
emissions and precession~\citep{Bonazzola:1995rb,Cutler:2002nw,2003MNRAS.341.1020W}. They also influence the thermal evolution of the
star~\citep{Vigano:2021olr,1996ApJ...473..322T}, among other effects.

The fate of NSs and possible gravitational and electromagnetic (EM) signals are therefore influenced by their internal magnetic field configuration,
though its geometry remains largely uncertain. Observations of pulsar spinning down suggest that the external magnetic field is predominantly poloidal
and dipolar, a configuration that aligns with the lighthouse-like radiation pattern typical of radio pulsars. However, this focus on the dipolar component
is a strong simplification~\citep{Ascenzi:2024wws}, as various internal magnetic field configurations could produce a similar external
appearance~\citep{Braithwaite:2005md,10.1111/j.1365-2966.2008.14034.x}, potentially masking the true complexity of the stellar interior.

The challenge of probing the internal magnetic field geometry has driven significant efforts into equilibrium models of magnetized NSs.
Early models focused on simple magnetic configurations, such as purely poloidal or purely toroidal
fields~\citep{1956ApJ...123..498P,1973MNRAS.161..365T,Kiuchi:2008ss,2013PhRvL.110g1101L}.
Though in recent years, numerical studies have increasingly explored more complex mixed poloidal-toroidal fields, motivated by the need
to more accurately represent the conditions within NSs (see e.g.~\cite{Ciolfi:2009bvAxisymmetric,10.1093/mnras/stu215,2015MNRAS.450.1638G,Tsokaros_2022,
  2025PhRvD.111f3030C} and references therein).
An open question still remains regarding the long-term stability of magnetic fields, as they must endure for timescales far exceeding
the dynamical ones. It is well known that purely poloidal or purely toroidal magnetic field configurations can indeed be equilibrium solutions.
However, such configurations tend to be dynamically unstable. Purely poloidal fields tend to become unstable due to the so-called pinch
instability~\citep{1954RSPSA.223..348K,10.1093/mnras/162.4.339,1999A&A...349..189S,Braithwaite:2005md}, while toroidal fields are prone to
the so-called Tayler instabilities~\citep{1973MNRAS.161..365T}, and of particular importance the kink instability~\citep{Begelman:1997gd,2009MNRAS.395.2162L,10.1111/j.1365-2966.2008.14034.x}.
These instabilities may also be triggered in more general magnetic field configurations.
The timescale for the growth of these instabilities is typically on the order of the Alfv\'en timescale 
\begin{equation}
  \tau_{\rm A} \sim  10~{\rm ms}  \left( \frac{R_{\rm NS}}{10~{\rm km}} \right) \left( \frac{B}{10^{15}~{\rm G}}
  \right)^{-1} \left( \frac{\rho}{10^{14}~{\rm g\, cm^{-3}}} \right)^{1/2}\,,
\end{equation}
$R_{\rm NS}$ is the typical length scale of the system (radius of the NS), $B$ is the magnetic field strength, and $\rho$ is the
rest-mass density. It is worth mentioning that usually the so-called Flowers-Ruderman mechanism~\citep{Florwers-ruderman}, an
oversimplified toy model, is invoked to explain the instability of the poloidal magnetic fields. It uses the analogy of two bar
magnets trying to align. Magnetic field lines can realign to lower the system's energy, driving an instability.
This process is influenced by the stellar rotation, with different timescales depending on the relation between the star rotation
period $T_{\rm r}$ and the Alfv\'en time $\tau_{\rm{A}}$. For $T_{\rm r} \gg \tau_{\rm{A}}$, the instability timescale is approximately
$\tau_{\rm{A}}$, while for $T_{\rm r}\ll \tau_{\rm{A}}$, it becomes $\tau_{\mathrm{MHD}}\approx 2\,\tau_{\rm{A}}^2/T_{\rm r}$,
the timescale for the magnetic field to realign through internal fluid motions in a rotating, conductive fluid. 

Recently magnetohydrodynamic (MHD) studies (see e.g.~\cite{Braithwaite_Nordlund,Ciolfi:2009bv,
2009MNRAS.395.2162L,Ciolfi:2010td,2013MNRAS.430.1811G,Tsokaros_2022, 2025PhRvD.111f3030C} and references 
therein) have suggested that NS magnetic fields are much more complex. These fields are likely a mixture
of poloidal and toroidal components, which interact in intricate ways that were not fully captured by this mechanism.
Mixed poloidal-toroidal configurations are often considered more stable and realistic for modeling the magnetic fields of
NSs~\citep{Braithwaite_Nordlund,Ciolfi:2009bv,2009MNRAS.395.2162L,Ciolfi:2010td,2013MNRAS.430.1811G,Tsokaros_2022, 2025PhRvD.111f3030C}.
The superposition of poloidal and toroidal magnetic field components may lead to a more stable configuration 
than configurations with only one component, as they can potentially counterbalance each other's instabilities.
However, the overall stability depends on the specific configuration and the relative strengths of the poloidal and toroidal components.
Numerical simulations in both Newtonian and general relativistic frameworks have shown that these configurations can remain
stable over significant timescales, though they may eventually succumb to instabilities, such as varicose or 
kink instabilities (see e.g. \cite{Tsokaros_2022}).
The latter instability is triggered when the toroidal magnetic field becomes too strong relative to the stabilizing 
forces, leading to displacements in the plasma perpendicular to the field 
lines~\citep{2011MNRAS.412.1730L,1973MNRAS.163...77M,10.1111/j.1365-2966.2008.14034.x}.

Over the past two decades, studies on the stability of the magnetic field of NS have increasingly 
focused on the field geometry, i.e. the distribution of the poloidal and toroidal components, rather 
than physical mechanisms that affect its evolution. In this work, we probe how the angular velocity 
of the NS affects the evolution of the magnetic field. In pursuit of this goal, we perform general 
relativistic magneto-hydrodynamic (GRMHD) simulations of rapidly rotating, self-consistent magnetized 
NSs with a pulsar-like magnetic field extending from the stellar interior into the exterior.
The stars that are initially uniformly rotating with angular velocities ranging from
non-rotation to half the mass-shedding limit.
The maximum magnetic field strength in our configuration reaches~$\sim 10^{17}\,\rm G$.
We note that astronomical observations primarily constrain the magnetic field strength at the surface 
of the NS (particularly at the poles) rather than the strength within the NS, which can be significantly higher.
The virial theorem provides an upper limit, suggesting that the internal magnetic field strength could reach up 
to $\sim 10^{18}\rm\,G$~\citep{1969ApJ...157..869G,reisenegger2013magneticfieldsneutronstars}.
Although this field strength is significantly larger from an astronomical perspective, it is dynamically weak, with
a magnetic-to-gas-pressure ratio $\beta^{-1}=P_{\mathrm{mag}}/P\ll 1$.
This ensures that we can effectively capture magnetic instabilities within the constraints of our computational resources.

We find that rotation can suppress or at least delay the onset of various magnetic instabilities for at least $10$ Alfv\'en times.
Unlike cases with low rotation, where the system loses about $99\%$ of its initial magnetic energy in only $4$ Alfv\'en times 
from the onset of the instability, highly rotating cases retain up to $30\%$ of their magnetic energy. 
Our results suggest that rotation may be crucial in stabilizing the magnetic field of NSs, regardless of its initial configuration.
We also observe that the star spontaneously develops differential rotation within its core followed by
a magnetic field misalignment with respect to the angular velocity.

The remainder of the paper is organized as follows.
In Sec.~\ref{sub_sec:sim_setup} we briefly review our numerical methods and their implementation, referring the reader to 
\citep{Cordero_Carri_n_2009,Bucciantini_2011,10.1093/mnras/stx1176,10.1093/mnras/stu215,10.1093/mnras/stu2628,Cheong_2021} 
for further details and code tests. A detailed description of the adopted initial data and the grid structure employed
for the evolution is given in Secs.~\ref{Sec:Initial_data} and~\ref{sec:gid}. 
A suite of diagnostics used to verify the reliability of our numerical calculations are summarized in Sec.~\ref{Sec:diagnostics}.
In Sec.~\ref{sec:Binstabilities}, we review the magnetic instabilities our configurations may be unstable to.
We present our results in Sec.~\ref{Sec:Results} and summarize our results and conclude in Sec.~\ref{Sec:conc}. Additionally,
in Appendices~\ref{sec:atm_study} and~\ref{sec:res_study} we probe the impact of the variable atmosphere present in our 
simulations and the impact of the resolution in our results. Throughout the paper we adopt Heaviside–Lorentz units, for which 
the speed of light $c$, gravitational constant $G$, solar mass $M_{\odot}$, vacuum permittivity $\epsilon_0$, are set to one 
($c=G=M_{\odot}=\epsilon_0=1$).
%
\section{Numerical Methods}
\label{Sec:Numerics}
\subsection{Simulation setup}
\label{sub_sec:sim_setup}
The numerical methods used in this study have been thoroughly detailed in previous works 
(see, e.g., \cite{Cheong_2020,Cheong_2021,Cheong_2022,Cheong_2023}). Here, we provide only 
a brief overview of the key aspects, directing the reader to those references for a detailed
discussion and code validation.

We perform full 3D ideal GRMHD simulations using the well-tested \texttt{Gmunu} code.
The code solves the Einstein equations using the approximation so-called extended conformal
flatness condition (xCFC)~\citep{Cordero_Carri_n_2009}, which equations are elliptical 
in nature, using a non-linear cell-centred multi-grid approach (see Eqs.~(72)-(76) in~\cite{Cheong_2021}). 
The matter and magnetic fields are evolved using the equations of ideal GRMHD, which are cast in a conservative scheme (see~Eqs.~(25)-(28)
in~\cite{Cheong_2021}), via a high-resolution shock capturing technique employing the 3rd-order reconstruction piecewise parabolic
method~\citep{1984JCoPh..54..174C} coupled to the Harten, Lax, and van Leer approximate Riemann solver~\citep{harten1983upstream}.
For time integration, strong-stability preserving Runge–Kutta of third order (SSPRK3)~\citep{1988JCoPh..77..439S} is used with a
Courant-Friedrichs-Lewy (CFL) factor set to 0.25.
To ensure the magnetic field remains divergenceless during the whole evolution, we integrate the magnetic induction equation using
the so-called staggered-meshed constrained transport~\citep{1988ApJ...332..659E}.
\texttt{Gmunu} uses adaptive mesh refinement (AMR) to resolve different scales accurately when needed. 
The parallelization and AMR of \texttt{Gmunu} is provided by the coupling \texttt{MPI-AMRVAC}
toolkit~\citep{Xia_2018,keppens2020mpiamrvacparallelgridadaptivepde,2023A&A...673A..66K}, which is an open-source message passing
interface (MPI) based parallelized toolkit with a pure block-tree AMR module.

Empirical results from~\cite{Bucciantini_2011,Cheong_2021} suggest that solving the xCFC equations every $10$ to $50$ time steps
provides an acceptable balance between accuracy and efficiency for isolated NS simulations.
In the simulations reported here, we choose to solve the xCFC equations every $10$ time steps to maintain consistency with these findings.
Finally, we adopt a $\Gamma-$law equation of state (EoS) $P = (\Gamma -1)\rho\,\epsilon$, allowing for
shock heating. Here, $\epsilon$ is the specific internal energy and $P$ is the pressure. We set $\Gamma=2$, which reduces to a polytropic law 
for the initial (cold) NS matter (see below). 
%
\subsection{Initial data}
\label{Sec:Initial_data}
We consider equilibrium configurations of uniformly rotating and magnetized NS, spanning from non-rotating 
to those approaching half the mass-shedding limit.
As our goal is to probe the effects of the angular velocity on the emergence of magnetic instabilities, we model the stars with a simple
polytropic EoS $P=K\rho_0^\Gamma$. The polytropic constant $K$ is set to $218.23\,\rm km^2$, what induces a Tolman-Oppenheimer-Volkoff
(TOV) maximum mass of $\sim 1.64\,M_\odot$.

The initial data (ID) of the axisymmetric equilibrium configurations are computed using the open source code {\tt Xtended Numerical Solver}
(\texttt{XNS})~\citep{Bucciantini_2011,10.1093/mnras/stx1176,10.1093/mnras/stu215,10.1093/mnras/stu2628}.
These models have a purely poloidal magnetic field defined through the magnetization function (see Eq.~(32) in \cite{10.1093/mnras/stu215})
\begin{equation}
     \mathcal{M}\left(A_{\phi}\right)=k_{\mathrm{pol}}\left(A_{\phi}+\xi\frac{1}{2}A_{\phi}\right),
 \end{equation}
which defines the distribution of the magnetic field (see left column in Fig.~\ref{fig:P1Ua_3D}).
Here, $A_{\phi}$ is the vector potential, $k_\mathrm{pol}$ is the poloidal magnetization constant that controls the strength of the magnetic
field, and $\xi$ is the non-linear poloidal term. We choose $k_{\mathrm{pol}}$ such that the maximum value of the magnetic field $B_\mathrm{max}
\approx 1.5 \times 10^{17}\,\rm G$ for all models listed in Table~\ref{tab:velocities}. While the resulting magnetic field strength is large
from an astrophysical point of view, it is dynamically weak (i.e. $\beta^{-1}\ll 1$) and enables us to resolve the magnetic instabilities with
the finite computational resources at our disposal.\footnote{The stronger the magnetic field, the shorter the growth timescale $\tau_{\rm A}$
of the instabilities.} We note that such strong magnetic fields appear in magnetar scenarios (i.e. NSs with extremely strong magnetic fields
$\gtrsim 10^{15}\,\rm G$) have been suggested as a likely source of Soft Gamma repeaters and Anomalous X-ray Pulsars~\citep{1996ApJ...473..322T,2008A&ARv..15..225M}.
In addition, during and immediately following core collapse and the formation of a proto-neutron star, the immense reservoir of free energy
can drive the amplification of magnetic fields to extreme strengths, potentially reaching up to $\sim 10^{17}-10^{18}\,\rm G$~\citep{Burrows:2007yx,2016MNRAS.460.3316R}.

We consider a set of five magnetized NS models, a non-rotating reference model named P1U0 (i.e. poloidal-uniform-number-case) and four rotating 
models~(see Table~\ref{tab:velocities}).
In all models we set the central rest-mass density of $\rho_{\rm c} \sim 7.9 \times 10^{14}\,\rm g\,cm^{-3}$.
The angular velocity ranges from slow rotation (P1U1 case) to half the mass-shedding limit (P1U4 case). 
%
%
\begin{table*}
\caption{Summary of the initial properties of the magnetized NS configurations.
    We list the name of the configuration P1UX (poloidal-uniform-number-case), the baryonic mass $M_0$, the rotational period $T_{\rm r}$ (ranging
    from zero to half of the mass-shedding limit), 
    the coordinate equatorial radius $R_\mathrm{e}$, the polar-to-equatorial-radius ratio $R_\mathrm{p}/R_\mathrm{e}$, 
    the total magnetic energy, the magnetic-to-binding-energy ratio $E_B/\mathcal{W}$, and the kinetic-to-binding-energy ratio $E_\mathrm{K}/\mathcal{W}$.
    In all of these models, the central rest-mass density is fixed to~$\rho_\mathrm{c} \simeq 7.9\times10^{14}\,\rm g\,cm^{-3}$.
    }
\begin{ruledtabular}
\begin{tabular}{c|cccccccc}
Case & $M_0\,[M_{\odot}]$ & $T_{\rm r}\,[\rm ms]$ & $R_\mathrm{e}\,[\rm km]$  & $R_\mathrm{p}/R_\mathrm{e}$ & $E_{{B_{\rm tot}0}}$[erg] &$E_{\mathrm{B}_{\rm tot}0}/\mathcal{W}$ &$E_\mathrm{K}/\mathcal{W}$ \\
\colrule
P1U0  & 1.51 & $\infty$ & 12.01  & 0.98  & $1.50\times10^{51}$   &$4.26\times 10^{-3}$ & 0.0   & \\
P1U1  & 1.52 & 6.19      & 12.01  & 0.98 & $1.55\times10^{51}$   &0.33                  & $3.16\times 10^{-3}$  &\\
P1U2  & 1.54 & 3.09      & 12.22  & 0.94 & $1.65\times10^{51}$   &0.33                  & $1.29\times 10^{-2}$  &\\
P1U3  & 1.59 & 2.06      & 12.56  & 0.89 & $1.66\times10^{51}$   &0.32                  & $2.95\times 10^{-2}$  &\\
P1U4  & 1.66 & 1.55      & 13.26  & 0.81 & $1.82\times10^{51}$   &0.30                  & $5.49\times 10^{-2}$  &
\end{tabular}
\end{ruledtabular}
\label{tab:velocities}
\end{table*} 

Following~\cite{Ruiz:2018wah,Paschalidis:2014qra}, at the beginning of the simulations, we impose a low-density, 
variable magnetosphere with a magnetic-to-gas pressure 
ratio of $\beta^{-1} \leq 10^{-3}$ in the exterior of the star.
This method has been successfully employed in previous studies (see e.g.~\cite{Ruiz:2021gsv}) to reliably evolve magnetic 
fields in environments dominated by magnetic pressure. Following~\cite{2025PhRvD.111f3030C}, we consider that any point at which 
the rest-mass density $\rho$ falls below the threshold $\rho_{\rm thr}$ is part of the atmosphere.
This threshold is set to be eleven orders of magnitude smaller than the initial maximum density, i.e. 
we set $\rho_{\rm thr} = 10^{-11} \, \rho_{\rm max}(t=0) \sim 4\times10^{-14} $.
For points falling below this threshold, the rest mass density is reset to $\rho_{\rm atm}=0.9\rho_{\rm thr}$.
Through empirical numerical experiments, we find that the non-rotating, slow and middle rotation models are stable
with a magnetic-to-gas-pressure ratio of $\beta^{-1}=10^{-6}$. However, higher rotation modes require a heavier atmosphere
likely due to larger magnetic gradients induced by rotation. For these cases, we find that 
the magnetic-to-gas pressure ratio should be
$\beta^{-1}=10^{-4},\,10^{-3}$ for P1U3 and P1U4, respectively. 
To assess the robustness of the numerical evolution 
of the magnetized NS against the effects of the artificial atmosphere treatment, we evolve model P1U2 using the three 
setups described above (see Appendix~\ref{sec:atm_study}). We find that this treatment has a minimal impact on the final outcome
of the magnetic field configuration, and so on the magnetic energy.
%
%
\subsection{Grid structure}
\label{sec:gid}
All NS configurations in Table~\ref{tab:velocities} are evolved using Cartesian coordinates $(x,y,z)$ without imposing any symmetries to prevent the 
suppression of some instabilities. The computational domain spans $[-180\,\rm km, 180\,\rm km]$ along each direction with a resolution of $N_x \times N_y 
\times N_z = 128^3$ grid points, and using four AMR levels differing in size and resolution by a factor of two.
The outer boundary is located at $\sim 15\, R_{\rm e}$, where $ R_{\rm e}$ is the coordinate equatorial stellar radius (see Table~\ref{tab:velocities}).
The highest-resolution grid covering the NS has a spatial resolution of~$\Delta x = \Delta y = \Delta z \approx 345~\rm m$. To probe the impact of the
numerical resolution on our results (see Appendix~\ref{sec:res_study}), we run the P1U2 model at two additional resolutions: a ``medium" resolution (MR), in which the 
finest refinement level has grid spacing of~$\Delta x = \Delta y = \Delta z \approx 230~\rm m$, and a ``high" resolution (HR), in which the finest level
has spacing~$\Delta x = \Delta y = \Delta z \approx 173~\rm m$.
Refinements are fixed after initialization, as a significant star expansion is not observed.

%
\subsection{Diagnostics}
\label{Sec:diagnostics}
To verify the reliability of our simulations, we monitor the conservation of the total mass $M_{\rm int}$ (see e.g. Eqs.~(3.128) and (3.191)) 
in~\citep{BaumgarteBook2010}, which coincides with the ADM mass $M_{\rm ADM}$ when evaluated at spatial infinity, as well as
the conservation of the rest-mass $M_{\rm rest}$ (see e.g. Eq.~(3.127) in~\citep{BaumgarteBook2010}, and the proper mass $M_{\rm proper}$ 
computed in the case of the xCFC spacetime as
\begin{eqnarray}
    M_{\rm ADM} &=& \displaystyle\int \left( \rho_H + \frac{K^{ij}K_{ij}}{16 \pi}\right) \psi^5 \dd{x}^3\,,
    \label{eq:Madm}\\
	M_{\rm rest} &=& \int \rho\, W\,\psi^6 \dd{x}^3, \label{eq:Mrest}\\
	M_{\rm proper} &=& \int \rho\,W\,\left( 1 + \varepsilon \right) \psi^6 \dd{x}^3,
        \label{eq:Mpr}
\end{eqnarray}
with $W$ the Lorentz factor, $\varepsilon$ the fluid specific internal energy, $\psi$ the conformal factor, $K^{ij}$ is the extrinsic curvature, and 
\begin{equation}
 \rho_H= T_{\alpha\beta}n^\alpha n^\beta =  \rho h W^2 - P + \frac{1}{2}(E^2+B^2)\,. 
\label{eq:rhoH}
\end{equation}
We note that the total stress-energy tensor $T_{\alpha\beta}$ is the sum of the stress-energy tensor for a perfect fluid and the stress-energy tensor
for the electromagnetic field, $n^\alpha$ is the normal to the hypersurface, $P = P_{\rm gas}$ is the pressure of the fluid, $E^2=E^iE_i$, $B^2=B^iB_i$,
and $E^i$, $B^i$ are the purely spatial electric, magnetic fields with respect to the normal observer, and $h=1+\varepsilon +P/\rho$ is the specific enthalpy.

\begin{figure}
\includegraphics[width=1.05\columnwidth]{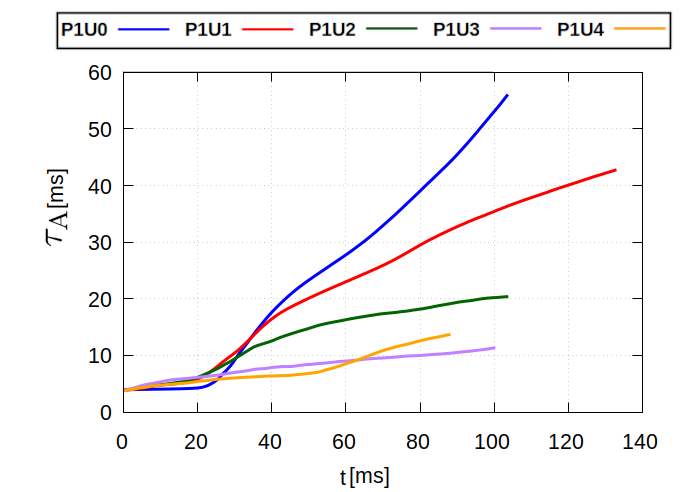}
\caption{Alfv\'en time $\tau_A$ as defined in Eq.~\eqref{eq:alf_time} versus the coordinate time 
for models in Table~\ref{tab:velocities}.}
\label{fig:TA_vs_ms}
\end{figure}

\begin{figure}%
\centering{
\includegraphics[width=0.47\textwidth]{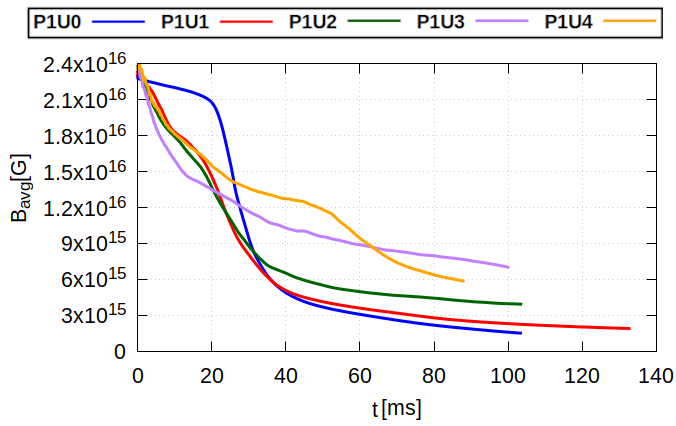}
\caption{Evolution of the average magnetic field strength $B_{\mathrm{avg}}$
vs. the coordinate time for models listed in Table~\ref{tab:velocities}. The 
rate at which the magnetic field decays strongly depends on the initial angular velocity
of the system.
 \label{fig:alfven}}}
\end{figure}

Consistent with~\cite{Cheong:2020kpv}, we find that in all cases listed in Table~\ref{tab:velocities} the above masses and the ADM angular momentum are
conserved to within $\lesssim 1.6\,\%$ throughout the entire evolution~\citep{Cordero_Carri_n_2009}. We calculate the internal energy, the kinetic energy,
the pressure contribution, and the electromagnetic energy defined as
\begin{eqnarray}
	E_{\rm int} &=& \int \epsilon_{\rm int}\, \psi^{6} \dd{x}^3\,, 
    \label{eq:E_int}\\
	E_{\rm kin} &=& \int \epsilon_{\rm kin}\,\psi^{6} \dd{x}^3\,, 
    \label{eq:E_kin}\\
	E_{\rm prs} &=& \int \epsilon_{\rm prs}\,\psi^{6} \dd{x}^3\,, 
    \label{eq:E_prs}\\
	E_{\rm EM}  &=& \int \epsilon_{\rm EM}\,\psi^{6} \dd{x}^3\,,
    \label{eq:E_EM}
\end{eqnarray}
where,
\begin{eqnarray}
	\epsilon_{\rm int} &=& \rho\, W^2\,\varepsilon\,, \\
	\epsilon_{\rm kin} &=& \rho\,W\,\left( W - 1 \right)\,, \\
	\epsilon_{\rm prs} &=& P\,\left( W^2 - 1 \right)\,, \\
	\epsilon_{\rm EM}  &=& \frac{1}{2}\left(B^2 + E^2\right)\,.
\end{eqnarray}
The total energy can be calculated either by adding the individual components $E_{\rm tot}=E_{\rm int} +E_{\rm kin} +E_{\rm prs} +E_{\rm EM}$
or by integrating the total conserved energy density over the entire volume as
\begin{equation}
	E_{\rm tot} = \int \mathcal{E}_{\rm tot}\,\psi^{6} \dd{x}^3,
 \label{eq:total_energy}
\end{equation}
with
\begin{equation}
 \mathcal{E}_{\rm tot} =  \rho\,h\,W^2 - P + \frac{1}{2}\,(E^2+B^2)  - \rho W . 
\label{eq:epsilon_tot}
\end{equation}
The characteristic timescale for magnetic field evolution is the Alfv\'en time,
\begin{equation}
\tau_\mathrm{A}=\frac{2\,R_\mathrm{NS}\sqrt{\rho_\mathrm{avg}}}{B_\mathrm{avg}}\,,
\label{eq:alf_time}
\end{equation}
where $\rho_\mathrm{avg}$ and $B_\mathrm{avg}$ 
are the average rest-mass density and magnetic field strength computed within the bulk of the NS.
Since these averages evolve over time, the Alfv\'en time is a dynamic quantity that changes throughout 
the evolution. Fig.~\ref{fig:TA_vs_ms} shows the evolution of $\tau_{\rm A}$ as a function of the 
coordinate time for the cases in Table \ref{tab:velocities}. We note that $\tau_{\rm A}$ increases only 
after the onset of magnetic instabilities (see below). Following this, and depending on the initial 
angular velocity of the NS, the Alfv\'en time either increases linearly with coordinate time, or reaches 
a plateau where it grows more slowly. To account for this time dependence, following~\citep{Sur_2022}, 
we define the Alfv\'en crossing time as
\begin{equation}
T_{\mathrm{A}}=\int_{0}^{t}\frac{1}{\tau_{\mathrm{A}}(t)} \dd t \,.
\label{eq:alf}
\end{equation}
%
%
%
\begin{figure*}
	\centering
	\begin{tabular}{@{}l@{\hspace{0.4mm}}c@{\hspace{0.4mm}}c@{\hspace{0.4mm}}c@{\hspace{0.4mm}}c@{\hspace{0.4mm}}c@{}}
		{} &
		\textbf{} & \textbf{} & \textbf{} & \textbf{} & \textbf{} \\
		\rotatebox{90}{\hspace{15mm}\bf{P1U0}} & 
		\includegraphics[width=0.195\textwidth]{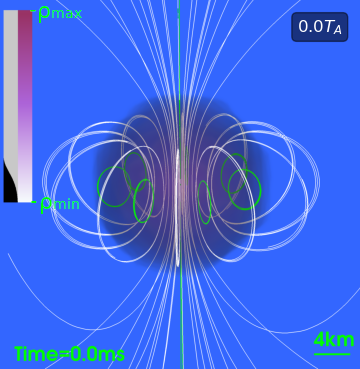} &
		\includegraphics[width=0.195\textwidth]{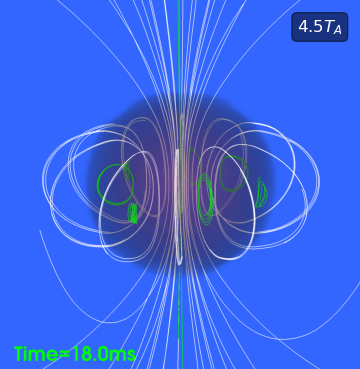} &
		\includegraphics[width=0.195\textwidth]{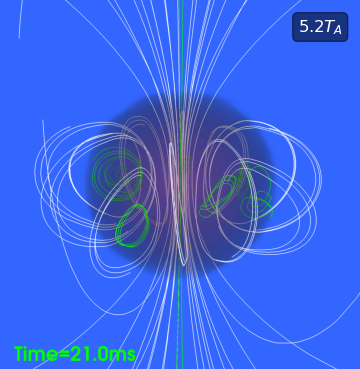} &
		\includegraphics[width=0.195\textwidth]{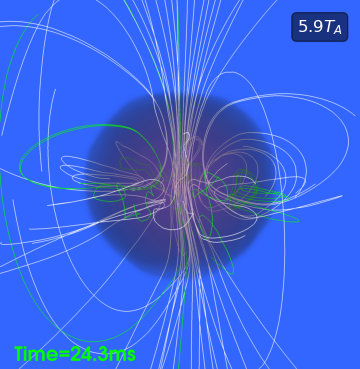} &
		\includegraphics[width=0.195\textwidth]{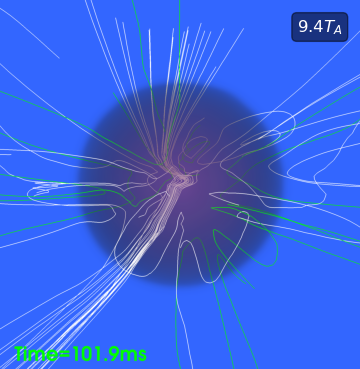} \\
		\rotatebox{90}{\hspace{15mm}\bf{P1U2}} & 
		\includegraphics[width=0.195\textwidth]{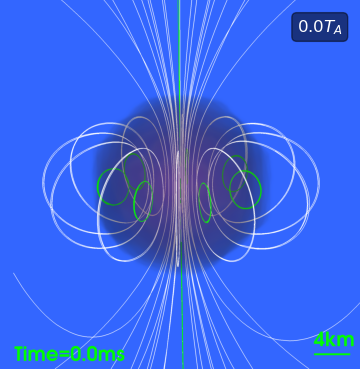} &
		\includegraphics[width=0.195\textwidth]{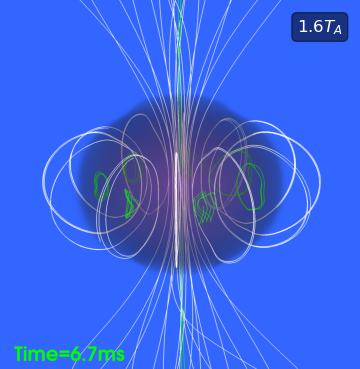} &
		\includegraphics[width=0.195\textwidth]{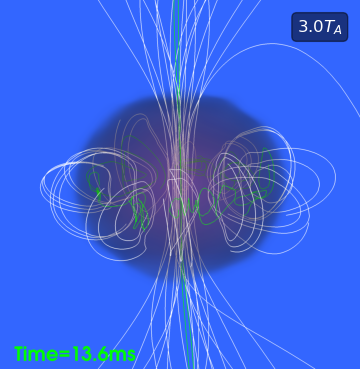} &
		\includegraphics[width=0.195\textwidth]{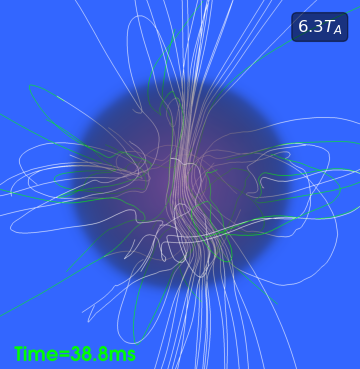} &
		\includegraphics[width=0.195\textwidth]{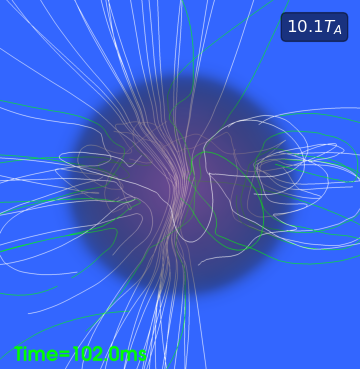} \\
		\rotatebox{90}{\hspace{15mm}\bf{P1U4}} & 
		\includegraphics[width=0.195\textwidth]{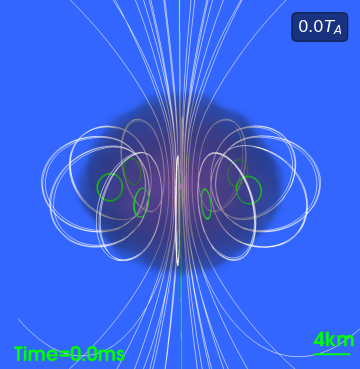} &
		\includegraphics[width=0.195\textwidth]{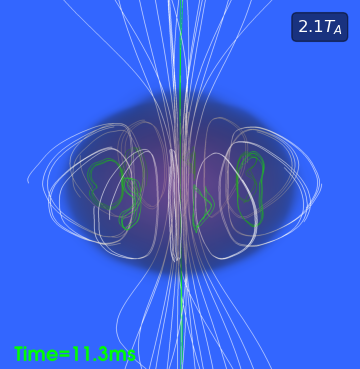} &
		\includegraphics[width=0.195\textwidth]{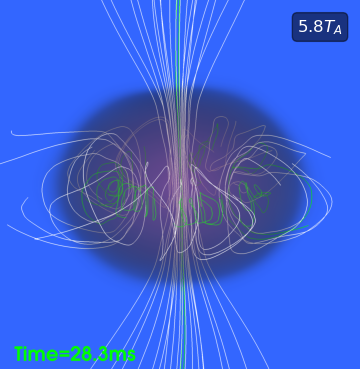} &
		\includegraphics[width=0.195\textwidth]{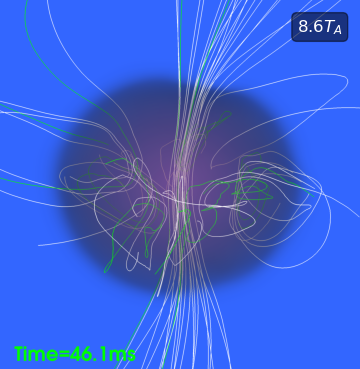} &
		\includegraphics[width=0.195\textwidth]{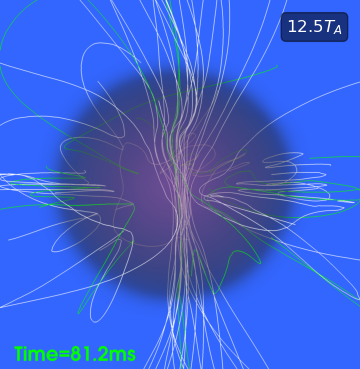} \\
	\end{tabular}	
	\caption{3D volume rendering of the rest-mass density $\rho$ at selected Alfv\'en crossing times
    for cases in Table~\ref{tab:velocities} along with the magnetic field lines. Green and white lines 
    display magnetic field lines that are either confined in the stars (green) or extend from the stellar interior 
    into the exterior (white).
	The colorbar (logarithmic scale) ranges from the atmospheric density to the maximum rest-mass density at the corresponding time. The black and gray rectangles to the left 
    of the colorbar indicate opacity, which ranges from fully opaque (black) to approximately $\sim 76\%$ opacity (gray).}
    \label{fig:P1Ua_3D}
        \label{fig:lines}
\end{figure*}
%
We rescale the evolution time using the Alfv\'en crossing period $T_{\mathrm{A}}$ and hence
a value of $T_{\mathrm{A}}=n$ indicates that the system has evolved for a time equivalent to $n\,\tau_{\rm{A}}$.
We note that $T_{\mathrm{A}}$ varies across different cases due to changes in the magnetic 
energy and the evolving density over time. 
During the beginning of the simulation, advancing by one $\tau_{\rm A}$ 
corresponds roughly to $4~\rm ms$ of physical time across all models. However, by the end of the simulations, a $\tau_{\rm A}$ 
corresponds to $\sim 55~\rm ms$ for model P1U0 and $\sim 20~\rm ms$ for P1U2. This difference arises because
the average magnetic field strength $B_{\mathrm{avg}}$ decays more rapidly in P1U0 compared to the other models, as shown 
in Fig.~\ref{fig:alfven}. Table~\ref{tab:sim_char} shows the evolution time in units of milliseconds and 
its corresponding Alfv\'en crossing time.

On the other hand, we define the poloidal and toroidal magnetic field components splitting $B^i$ into
\begin{equation}
B^i\equiv B_{\textup{tor}}\,\hat{e}^i_{\textup{tor}} + B_{\textup{pol}}\,\hat{e}^i_{\textup{pol}}\,, 
\label{eq:decomposition}
\end{equation}
where~$\hat{e}^i_{\textup{tor}}$, $\hat{e}^i_{\textup{pol}}$ are orthonormal 3-vectors such 
that~$q_{ij}\hat{e}^i_{\textup{tor}}\hat{e}^j_{\textup{tor}} = q_{ij}\hat{e}^i_{\textup{pol}}\hat{e}^j_{\textup{pol}} = 1$, 
$q_{ij}\hat{e}^i_{\textup{tor}}\hat{e}^j_{\textup{pol}} = 0$ and $\hat{e}^i_{\textup{tor}} \propto (-y,x,0)$ in 
Cartesian coordinates. Here $q_{ij}$ is defined as~\citep{Bamber:2024qzi}
\begin{align}
q_{ij} =& \gamma_{ij}(1 + w_k w^k) - w_i w_j\,, \\
w^i =& (v^i + \beta^i)/\alpha\,,
\end{align}
with $v^i$ the three velocity, and $\alpha$ and $\beta^i$ the lapse and the shift vector, respectively. 
We also define the magnetic field strength
\begin{equation}
  B=\psi^2 \sqrt{\delta_{i j} B^i B^j}\,,
\label{eq:strength}
\end{equation}
with $\delta_{i j}=\operatorname{diag}(1,1,1)$. 
\par
In addition, we monitor the total magnetic field energy
\begin{equation}
  E_{\mathrm{B_\mathrm{tot}}}=\frac{1}{2}\,\int B^i B_i\,\psi^{6} \dd{x}^3\,,
\label{eq:Ebtot}
\end{equation}
the toroidal magnetic energy,
\begin{equation}
   E_{\mathrm{B_\mathrm{tor}}}=\frac{1}{2}\int {(B_{\mathrm{tor}}})^i\,({B_{\mathrm{tor}}})_i\, \psi^{6} \dd{x}^3\,,
\label{eq:Ebtor}
\end{equation}
and the poloidal magnetic energy,
\begin{equation}
   E_{\mathrm{B_\mathrm{pol}}}=E_{\mathrm{B_\mathrm{tot}}}-E_{\mathrm{B_\mathrm{tor}}}\,,
\label{eq:Ebpol}
\end{equation}
to study how the toroidal and poloidal magnetic energy develops during the evolution.
As the system evolves, the magnitude and the geometry of the initial poloidal magnetic field are expected to change
due to the instabilities (see~Sec.~\ref{sec:Binstabilities}) in a timescale of the order of an 
Alfv\'en time, what has been confirmed by previous numerical studies~(see~e.g.~\cite{Braithwaite_2007,Sur_2022,Tsokaros_2022}).
We note that the initial value of the Alfv\'en time is $\sim 4~{\rm ms}$ for all models. 
%
%
\begin{table}
    \caption{Evolution times for all NS cases listed in Table~\ref{tab:velocities}.
    We list the time in units of milliseconds and Alfv\'en crossing time $T_{\rm A}$.
}
    \centering
    \begin{tabular}{c|cc}
\hline
\hline
       Model & Time [ms] & Time [$T_{\mathrm{A}}$] \\
       \colrule
       P1U0 & $103.70$ & $9.44$  \\
       P1U1 & $133.00$ & $9.58$\\
       P1U2 & $103.80$ & $10.16$  \\
       P1U3 & $100.30$ & $13.18$   \\
       P1U4 & $88.20$ & $13.04$ \\
\hline
\hline  
    \end{tabular}
    \label{tab:sim_char}
\end{table}

It is worth noting that rotation may change the timescale of certain instabilities. \cite{Pitts} showed that in 
rapidly rotating models, magnetic instabilities appear roughly at around $\tau_\mathrm{A}^2/T_{\rm r}$, where $T_{\rm r}$ is the stellar 
rotational period (see also \cite{Spruit:2007bt} and \cite{Florwers-ruderman}). Therefore, in our fastest rotating 
model (P1U4), we will expect the instabilities to be triggered at $\sim 10\,\rm ms$.
%
%
\section{Instabilities in magnetic fields and the role of rotation}
\label{sec:Binstabilities}
As discussed in Sec.~\ref{Sec:Intro}, magnetized NSs are prone to various 
magnetic instabilities that can considerably influence the evolution of their magnetic fields, 
potentially causing magnetic field reconfigurations and releasing energy in the form of heat, 
EM radiation, or even particle acceleration. These instabilities may contribute to the gradual decay of the magnetic field.
In the following, we review some of the most significant of these instabilities:

\begin{enumerate}
\item{\bf Tayler instability:}
  The instability occurs in systems where the magnetic field has a strong azimuthal (toroidal) component, commonly found in cylindrical or
  toroidal geometries. This instability is triggered when the magnetic energy stored in the azimuthal field becomes strong enough to
  destabilize the system, overcoming the stabilizing effects of fluid pressure and the tension in the magnetic field lines~\citep{1973MNRAS.161..365T}.
  It typically develops on the order of the Alfv\'en timescale $\tau_{\rm A}$. While our configurations initially have only a poloidal magnetic field
  component, a strong toroidal component emerges after a few dynamical timescales (see below). Two important modes of this instability are the so-called
  ``varicose'' (or ``sausage'') and the  ``kink''. These are characteristic deformations of the flux tubes. The former is related to radial contractions
  of the flux tubes, while the latter is associated with their twisting.
\item{\bf Pinch instability:}
  The instability is driven by an axial current, and typically appears in poloidal magnetic field configurations, which creates a circular magnetic field
  around the current. The magnetic pressure generated by the current flowing through the plasma is strong enough to compress the plasma inward.
If the inward magnetic pressure exceeds the outward pressure forces (thermal pressure and magnetic tension), the plasma becomes unstable, leading to 
collapse or constriction of the plasma column~\citep{1973MNRAS.162..339W,1986RvMP...58..741T}. The timescale for the pinch instability is also on the
order of the Alfv\'en timescale, similar to the Tayler instability. Also, like the Tayler instability, the pinch instability exhibits analogous varicose
and kink modes.
\item {\bf Parker or magnetic buoyancy instability:}
It occurs in systems where the magnetic field has a significant horizontal component, as in poloidal and mixed (poloidal + toroidal) configurations.
It is typically triggered in stratified (layering) environments where there are gradients in density or pressure.
This configuration is common in the crust of NSs, where the magnetic field and the stratification of matter are aligned.
The instability is triggered when buoyant forces act on the magnetic field lines, causing them to rise or fall~\citep{Parker,Acheson}.
It appears mainly if the magnetic field strength increases in the direction of gravity, with magnetic pressure counteracting gravitational forces.
Energy can be released as the magnetic field lines bend due to strong pressure gradients.
This instability generally requires stronger magnetic fields than the Tayler instability and may play a lesser role in certain contexts.
The timescale for the Parker instability is on the order of the Alfv\'en timescale.
\end{enumerate}

The MHD simulations of NSs endowed with a purely poloidal magnetic field in~\citep{Geppert2006MagnetarsVR} 
found that oblique configurations, with magnetic inclination $\lesssim 45^\circ$ could maintain high magnetic 
fields, while slower rotation or larger inclinations led to significant magnetic energy loss.
However, the simulations in~\citep{Braithwaite_2006,Braithwaite_2007} showed that this field configuration is 
unstable and decay within a few Alfv\'en times regardless of the angular velocity of the star, even in oblique 
configurations, although the rotation slows this decay.
In highly rotating stars, the instability timescale is lengthened by $t_{\rm A}/T$.
\cite{Kiuchi_2008} performed axisymmetric GRMHD simulations of NSs with a pure toroidal magnetic field. 
They found that rapid rotation may suppress the Tayler instability, suggesting that when kinetic energy is 
about six times larger than the electromagnetic energy, the magnetic field can remain stable.
However, axisymmetry may artificially suppress some instabilities~(see e.g.~\citep{ Acheson,Parker}). 
Subsequent 3D simulations by~\cite{Kiuchi_2011} of highly rotating NSs with toroidal magnetic fields suggested that magnetic 
instabilities, such as the Parker and Tayler instabilities, can indeed be triggered on the order of the Alfv\'en timescale.
The GRMHD studies by~\cite{Sur_2022} and~\cite{Tsokaros_2022} found that mixed toroidal and poloidal magnetic field 
configurations are unstable. More recently, \cite{2025PhRvD.111f3030C} probed the effect of resistivity on the emergence of 
magnetic instabilities in isolated NS endowed with both poloidal and toroidal magnetic field configurations. They 
found that although all configurations are unstable to the Tayler instability, resistivity can delay its growth.
%
\section{Results}
\label{Sec:Results}
The outcomes and basic dynamics for all our cases in~Table~\ref{tab:velocities} are similar, hence we 
show snapshots and discuss the evolution only for nonrotating (P1U0), middle (P1U2) and highest (P1U4) 
rotation cases, unless otherwise specified. We summarize key results for all cases in Table~\ref{tab:energt_ratio}.
\subsection{Magnetic field evolution}
\label{topo}
Fig.~\ref{fig:lines} displays the evolution of magnetic field lines for P1U0, P1U2, and P1U4 cases at different Alfv\'en 
crossing times. Note that, since these times are case-dependent (see Eq~\eqref{eq:alf} and Table~\ref{tab:energt_ratio}), 
we choose the times to highlight the onset and evolution of the magnetic instabilities for each individual case.
%
\begin{figure*}
\includegraphics[width=2.1\columnwidth]{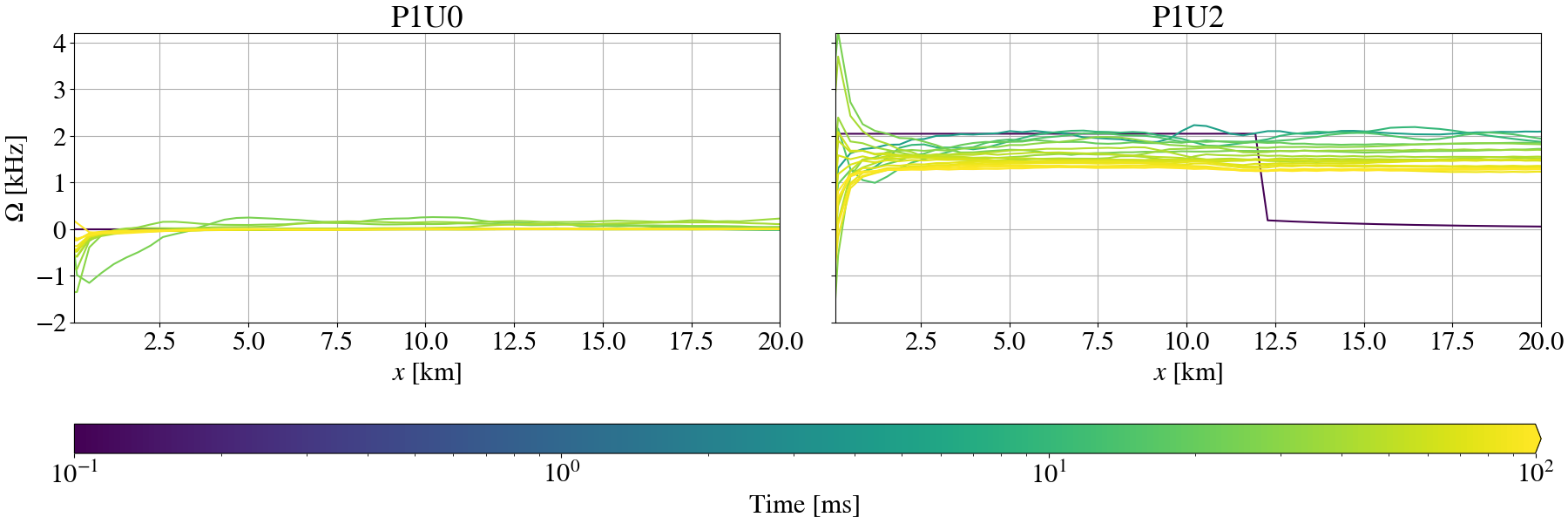}
\caption{Angular velocity profiles along the coordinate x-axis for the non-rotating case P1U0 (left) and the mildly rotating case P1U2 (right). 
Differential rotation spontaneously develops in the cores of the stars (roughly in the region contained within $\lesssim 3\,\rm km$) 
regardless of their initial rotation profile. Similar behavior is observed in the other cases in Table~\ref{tab:velocities}.}
\label{fig:omega_diff}
\end{figure*}

During the first $t\sim 3T_{\rm A}$ (or $\sim10\,\rm ms$), the pure poloidal magnetic field configuration of P1U0 
(see top-left panel in Fig.~\ref{fig:lines}) remains constant. Following this early epoch, the inner regions of 
the NS ($\lesssim 3\,\rm km)$ spontaneously develop differential rotation with a peak value of~$\sim 1~\rm kHz$ 
while the outer layers rotate roughly uniformly with $\Omega\sim 100\,\rm Hz$ (see left panel in Fig.~\ref{fig:omega_diff}).
We note that spontaneous differential rotation has also been observed in full GRMHD simulations of self-consistent 
rotating NSs seeded with poloidal and toroidal magnetic fields in~\citep{Tsokaros_2022}, and in resistive GRMHD 
simulations~\citep{2025PhRvD.111f3030C}. This effect has been attributed to the initial strong poloidal magnetic field component. 
The onset of differential rotation leads to the development of magnetic winding 
within the next $t\lesssim 1.5T_{\rm A}$, and so to the emergence of a strong toroidal magnetic field component. This 
new magnetized NS configuration is prone to instabilities that distort the magnetic 
field (see top row in Fig.~\ref{fig:P1Ua_3D}). In particular, we observe that by 
$t\sim 5T_{\rm A}$ the poloidal
field lines confined inside the NS (shown in green in the second and third panel in Fig.~\ref{fig:lines}) undergo a 
change in their cross-sectional area, a behavior consistent with the varicose mode, which can develop due to the pinch 
instability~\citep{10.1093/mnras/162.4.339}. This behavior has also been reported in~\citep{Sur_2022}. Simultaneously, 
and consistent with the Parker instability, weaker-tension field lines (mainly smaller loops) become buoyant due to local 
magnetic pressure, causing them to rise toward the lower-density regions of the star. As angular momentum is transported in 
the inner regions of the star, the poloidal magnetic field anchored in these regions is twisted into a toroidal configuration, 
causing the field lines to wind along the equatorial plane (third and fourth panels in Fig.~\ref{fig:lines}). We also observe 
that some field lines begin to shift slightly along the magnetic axis~\citep{Braithwaite_2007}, a footprint of the kink 
instability, likely associated with strong toroidal magnetic fields~\citep{1973MNRAS.161..365T}. By $t\sim 9T_{\rm A}$, 
the original magnetic field configuration has been completely disrupted due to the cumulative effects of the above magnetic
instabilities (see top-right panel). We note that once differential rotation is present, additional instabilities, such as 
the magnetorotational instability (MRI) may also operate within the star. However, given our current numerical resolution 
(see~Sec.~\ref{sec:gid}), it is unlikely that the inner regions of the star become unstable against 
the MRI~\citep{Kiuchi:2023obe}.

%

\begin{table} 
    \centering
     \caption{summary of the results. We list the fraction of the initial total magnetic energy $E_{B_{\rm tot}0}$
       remaining at~$T_{\rm A}-T\simeq 4$ (see discussion in the main text) following the onset of the magnetic instabilities, the
       toroidal-to-total magnetic energy ratio at that point, its maximum, and the average magnetic field tilt angle with respect to
       the rotation axis for all NS models in Table~\ref{tab:velocities}.
    \label{tab:energt_ratio}}
    \scalebox{0.95}{%
    \begin{tabular}{c|ccccccccc}
    \hline\hline
     Model & $E_{B_{\rm tot}}/E_{B_{\rm tot}0}$ & ${E_{{B_{\rm tor}}}}/{E_{{B_{\rm tot}}}}$ & max(${E_{{B_{\rm tor}}}}/{E_{{B_{\rm tot}}}}$) & $\zeta_{\rm avg}$ [º]\\
\colrule
        P1U0 &  $1.13\%$   &  $6.3\%$  & $10.5\%$ &  $-$ \\
        P1U1 &  $7.48\%$   &  $7.0\%$  & $14.7\%$ & $7.7$  \\
        P1U2 &  $16.79\%$  &  $12.3\%$ & $21.4\%$ & $14.9$ \\
        P1U3 &  $59.09\%$  &  $11.1\%$ & $12.8\%$ & $14.5$ \\
        P1U4 &  $19.12\%$  &  $14.1\%$ & $18.5\%$ & $19.6$ \\
\hline
\hline
    \end{tabular}}
\vspace{7pt}    
\end{table}

\begin{figure}
\includegraphics[width=0.235\textwidth]{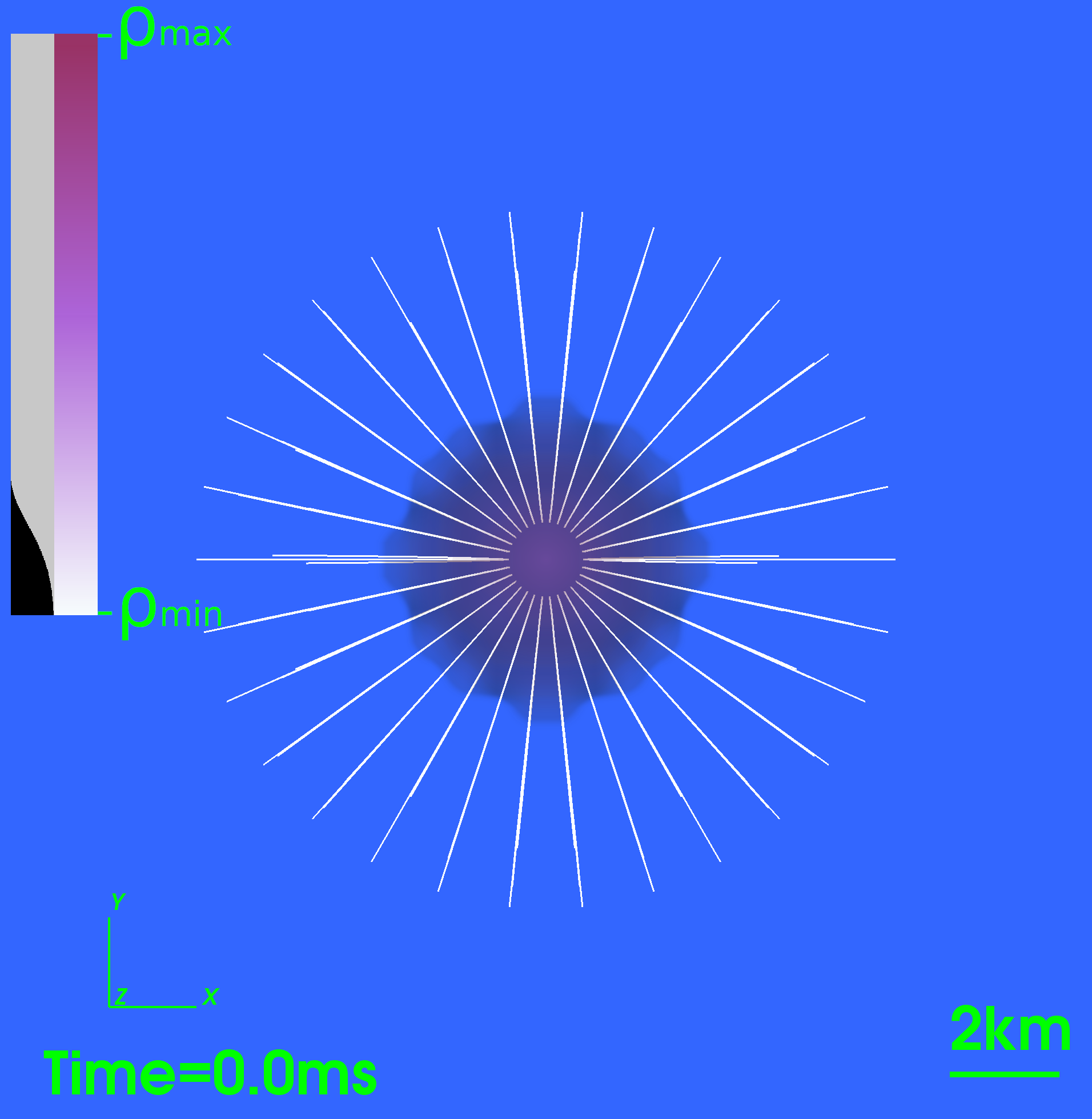} 
\includegraphics[width=0.235\textwidth]{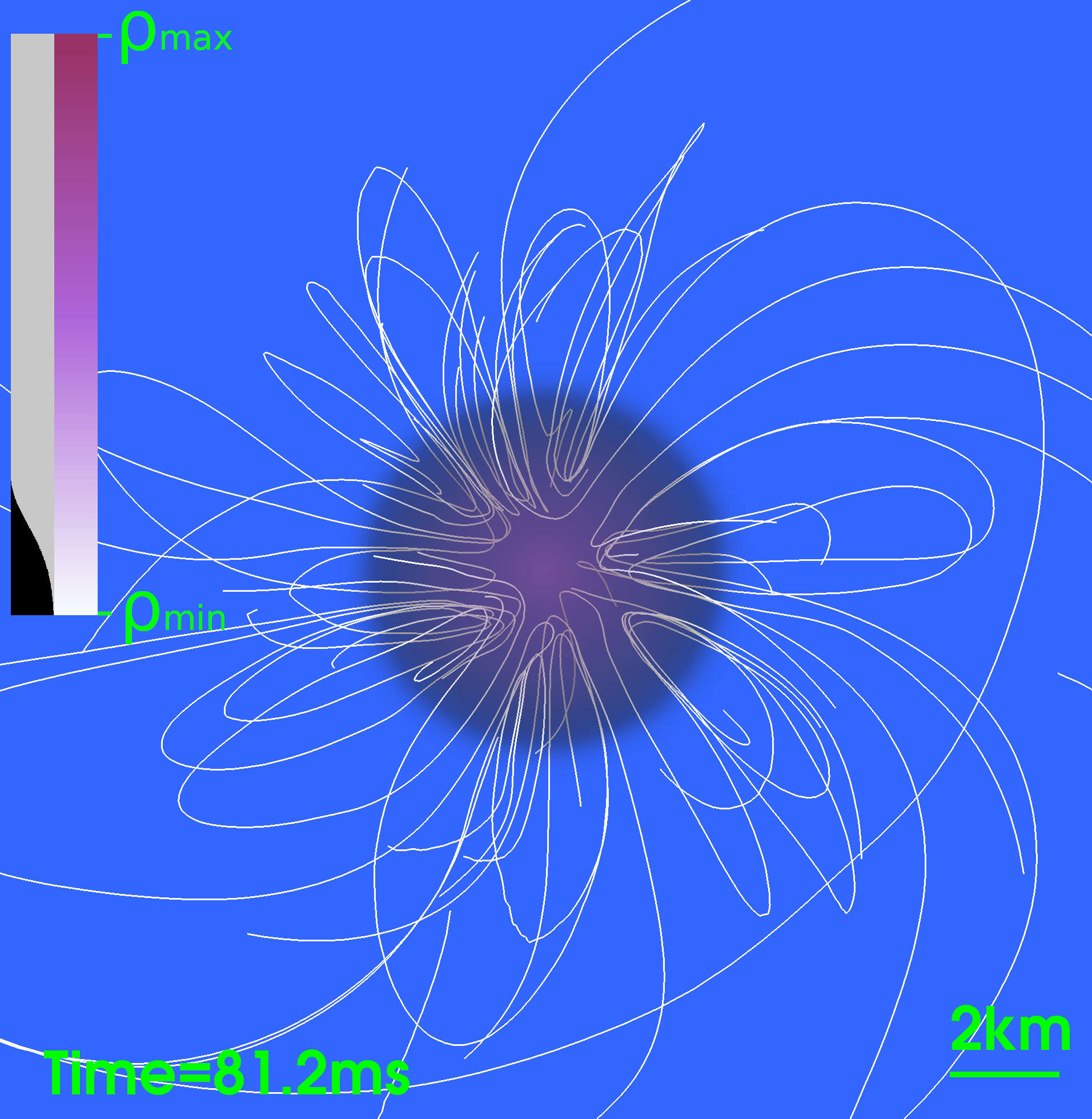}
\caption{Top view of the rest-mass density of P1U4, along with the field lines 
at $T_{\rm A}=0$ (left) and near to the end of the simulation ($\approx 12.5T_{\mathrm{A}}$).
The colorbar (logarithmic scale) ranges from the atmospheric density to the maximum rest-mass density at the corresponding time.
The black and gray rectangles to the left  of the colorbar indicate opacity, which ranges from fully opaque (black) to~$\sim 76\%$ opacity (gray).}
\label{fig:linesxy}
\end{figure}

In the uniformly rotating cases, we observe that during the early stages of evolution, the initial 
poloidal magnetic field configuration remains unchanged. Subsequently, differential rotation 
spontaneously develops on a shorter timescale than in P1U0 (see second column in Fig.~\ref{fig:lines} and Fig.~\ref{fig:omega_diff}), around
$t\sim 1.6T_{\rm A}$. Differential rotation winds the poloidal magnetic field into a toroidal 
configuration. The varicose mode is subsequently triggered in the stars, leading to observable changes 
in the cross-sectional area of the interior magnetic field lines (see second column in Fig.~\ref{fig:lines}).
In contrast to P1U0, where the Parker instability emerges due to magnetic pressure gradients, we find no
evidence of its development in these cases. It is likely that the additional centrifugal support 
reduces the magnetic pressure gradients within the star, thereby suppressing the Parker instability.
We note that local changes in the magnetic field and enhanced magnetic tension may also contribute 
to stabilizing the system against magnetic buoyancy modes~\citep{Ruiz:2021qmm,Aguilera-Miret:2024cor}. 
By~$t\gtrsim 6.0T_{\rm A}$, consistent with the onset of the kink instability, we observe slight 
displacements of some field lines along the magnetic axis (see the third and fourth panels in the 
middle and bottom rows of Fig.~\ref{fig:lines}). Finally, by $t\gtrsim 10T_{\rm A}$ magnetic instabilities 
have fully disrupted the initial magnetic field configuration (see fifth panels in the middle and bottom 
rows in Fig.~\ref{fig:lines}). Further evidence of the kink instability can be seen through a top view of 
P1U4 displayed in Fig.~\ref{fig:linesxy}. Initially, the field lines of the poloidal magnetic field 
(closed loops) are aligned along the rotation axis (left panel), but they become twisted at later times 
(right panel). A distorted quasi-axisymmetric ``flower-like" structure forms as a result of the growth of the toroidal 
component, consistent with the development of the kink instability. Similar behavior is observed in the 
other rotating cases in Table~\ref{tab:energt_ratio}. 
%
%
%
%
\subsection{Magnetic energy evolution}
\label{magenergy}
We now analyze the evolution of the models from an energetic point of view. We examine the total magnetic energy 
$E_{B_{\rm tot}}$ as defined in Eq.~\eqref{eq:Ebtot}, along with the toroidal and poloidal energies as 
given by Eqs.~\eqref{eq:Ebtor} and \eqref{eq:Ebpol}, respectively. We focus on assessing the stability of the 
systems and evaluating how rotation influences the emergence of a toroidal component and how both rotation and magnetic
instabilities contribute to its evolution. 

\begin{figure}
\includegraphics[width=0.47\textwidth]{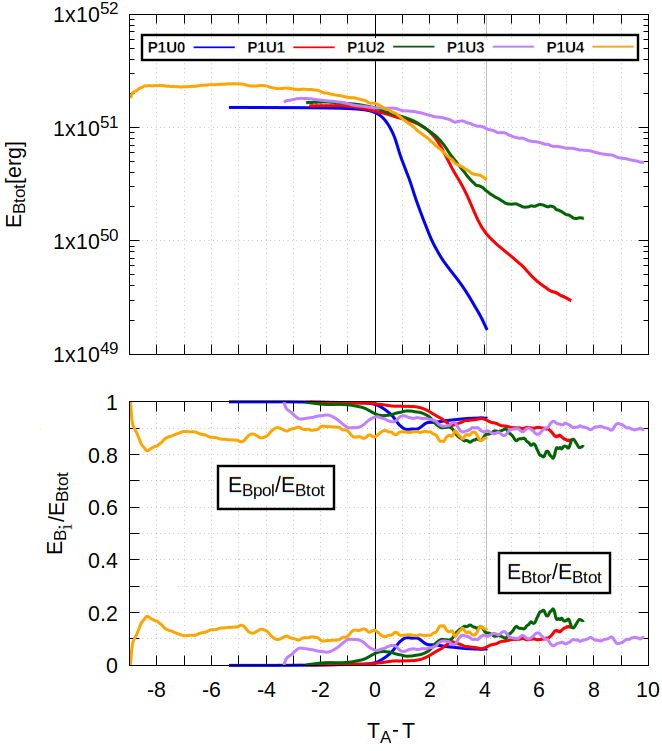}
\caption{Total magnetic field energy $E_{B_{\rm tot}}$ (up) and normalized toroidal magnetic 
field energy $E_{B_{\rm tor}}$ and poloidal magnetic field energy $E_{B_{\rm pol}}$ (down) vs. $T_{\rm A}-T$. 
Here $T$ is the time at which $E_{B_{\rm tot}}$ has decreased by $10\%$ of its initial value,
which is our definition of the onset of the magnetic instabilities. The vertical gray line marks the termination 
of the P1U0 and P1U4 simulations. $E_{\rm B_{i}}$ denotes either $E_{\rm B_{tor}}$ or $E_{\rm B_{\rm pol}}$ components.
\label{fig:casecombo2}}
\end{figure}

The upper panel of Fig.~\ref{fig:casecombo2} displays the total magnetic energy $E_{B_{\rm tot}}$ vs. the Alfv\'en crossing 
time for all cases and shifted by a time $T$. We define $T$ as the time 
at which the total magnetic energy has decreased by $10\%$. Therefore, negative times (~$t-T<0$) indicate how long the 
system remains stable before the instability sets in, while positive times ($t-T>0$) show how long the instabilities 
have been operating in the system.
 We note that before $t\sim 5\,T_{\rm A}$, the Parker and varicose instabilities are operating in P1U0 (see discussion above). 
However, the total magnetic energy remains roughly constant during this epoch. By $t\sim 5\,T_{\rm A}$, the magnetic energy 
begins to decline, marking the onset of the kink instability. By $t\sim 6\,T_{\rm A}$ (or $T_{A}-T\sim 0.5$), the initial magnetic
field configuration has been disrupted, and the magnetic energy drops sharply.

Although we evolve all cases for approximately~$100~\rm ms$ (see Fig.~\ref{fig:TA_vs_ms}), the onset of magnetic instabilities
limits the evolution of the P1U0 and P1U4 models to about $T_{\rm A}-T\simeq 4$, which is marked as a gray line in Fig.~\ref{fig:casecombo2}. 
To ensure a fair comparison across all models, we restrict our analysis to this period of time~and~then comment on the subsequent 
evolution for models that extend beyond this time.

We note that the slope decay of $E_{B_{\rm tot}}$ depends on the angular velocity. In general, the slower the star rotates, 
the steeper the slope becomes. The exception to this trend is P1U4, which we will analyze in Sec.~\ref{sec:roteffetc}. Table
\ref{tab:energt_ratio} shows the ratio between the total initial magnetic energy and the energies at $T_{A}-T=4$.
For the non-rotating model, the remaining energy is only $1\%$, while P1U3 retains up to $60\%$ (see top panel in
Fig.~\ref{fig:casecombo2}). This suggests that rotation may suppress or at least delay the onset of the instability, 
leading to a slower decline in magnetic energy. These results are in agreement with previous findings~\citep{Braithwaite_2007}.
However, we note that in P1U3 case, the most magnetically-stable case, the magnetic energy has not reached a steady equilibrium.
Toward the end of our simulation ($T_{\rm A}-T\sim 10$), the magnetic energy continues to decrease at the same rate.
We cannot determine whether it will eventually settle or continue to decline. We speculate that because of the presence of 
numerical dissipation, the energy will keep falling. To assess the stability of these new magnetic field configurations 
(such as the one in the right panel of Fig.~\ref{fig:linesxy}), higher-resolution and longer numerical simulations 
than those reported here are required to confirm this behavior. The energy values reported in Table \ref{tab:energt_ratio} 
should be taken as estimates, as they may differ with resolution (see Appendix~\ref{sec:res_study}).

The lower panel of Fig.~\ref{fig:casecombo2} displays both the poloidal $E_{B_{\rm pol}}$ and the toroidal 
$E_{B_{\rm tor}}$ energies. As expected, the latest is initially null. Although all our configurations are 
initially uniformly rotating, the systems spontaneously developed differential rotation (see Fig.~\ref{fig:omega_diff}) 
which induces the growth of the toroidal energy. In the non-rotating case the $E_{B_{\rm tor}}$ is $\sim 6\%$ of 
the initial magnetic energy, while in the highest rotating case the toroidal energy reaches a value of $\sim 14\%$ 
(see Table~\ref{tab:energt_ratio}). For model P1U0, the growth of $E_{B_{\rm tor}}$ coincides with the onset of the kink 
instability, i.e., the point where the total magnetic energy $E_{B_{\rm tot}}$ changes slope. We note that the toroidal 
energy peaks at about $10\%$ of $E_{B_{\rm tot}}$, corresponding to the first appearance of a non-zero toroidal magnetic 
field component at $T_{\rm A}-T\simeq 0.6$, as shown in Fig.~\ref{fig:P1Ua_3D}. Afterward, it decays to roughly $6\%$.
Before this epoch, the varicose instability mode appears earlier (see discussion above) at $T_{\rm A}-T\sim - 0.9$. This 
suggests that the varicose mode is likely associated with pinch-type instabilities related to the poloidal magnetic field 
configuration~\citep{10.1093/mnras/162.4.339}.

On the other hand, the overall behavior of the rotating models is basically the same, though the evolution of $E_{B_{\rm tot}}$ 
is more complex due to stellar rotation. Table~\ref{tab:energt_ratio} displays the toroidal magnetic energy at $T_{\rm A}-T=4$ 
for these models. Changes in the slope of the decay of $E_{B_{\rm tot}}$ appear to coincide with the growth of the toroidal 
energy component in P1U1 and P1U2, suggesting a link between toroidal field growth and energy decay. However, this correlation 
is not observed in P1U3 and P1U4, and is less clear in P1U1 and P1U2 compared to the non-rotating case. It is likely that rotation 
suppresses, or at least delay, the onset of the kink instability. This implies that even when $E_{B_{\rm tor}}$ increases, 
the total magnetic energy $E_{B_{\rm tot}}$ may remain, at least initially, largely unaffected.

The spontaneous emergence of a toroidal magnetic field component from a purely poloidal configuration has also been reported 
in~\cite{Sur_2022} and~\cite{cook2023grathena}. \cite{Sur_2022} found that the toroidal magnetic field energy decays to $1\%$ of the 
total magnetic energy at $\sim 6.5$ Alfv\'en times, reaching only a quasi-stable equilibrium. In this configuration, most of 
the magnetic energy is converted into heat, while the rest is radiated away via a Poynting flux. On the other hand, the final 
outcome of the non-rotating model in~\citep{cook2023grathena} corresponds to a configuration with a toroidal magnetic field 
accounting $10\%$ of the total magnetic energy after $\sim 44\rm ms$.
Regarding the decay of magnetic energy, this effect has also been observed in numerical simulations~(see e.g.~\cite{Braithwaite_2007,Sur_2022}).
\cite{Braithwaite_2007} found that rotation slows the decay. In their case, the fastest rotating model (at the mass-shedding limit) 
retains $\lesssim 25\%$ of the initial energy. However, a comparison between their results and ours is difficult, as they do 
not report the simulation time in Alfv\'en units nor do they specify how the magnetic energy was calculated.

\begin{figure}[h]
\includegraphics[width=1.06\columnwidth]{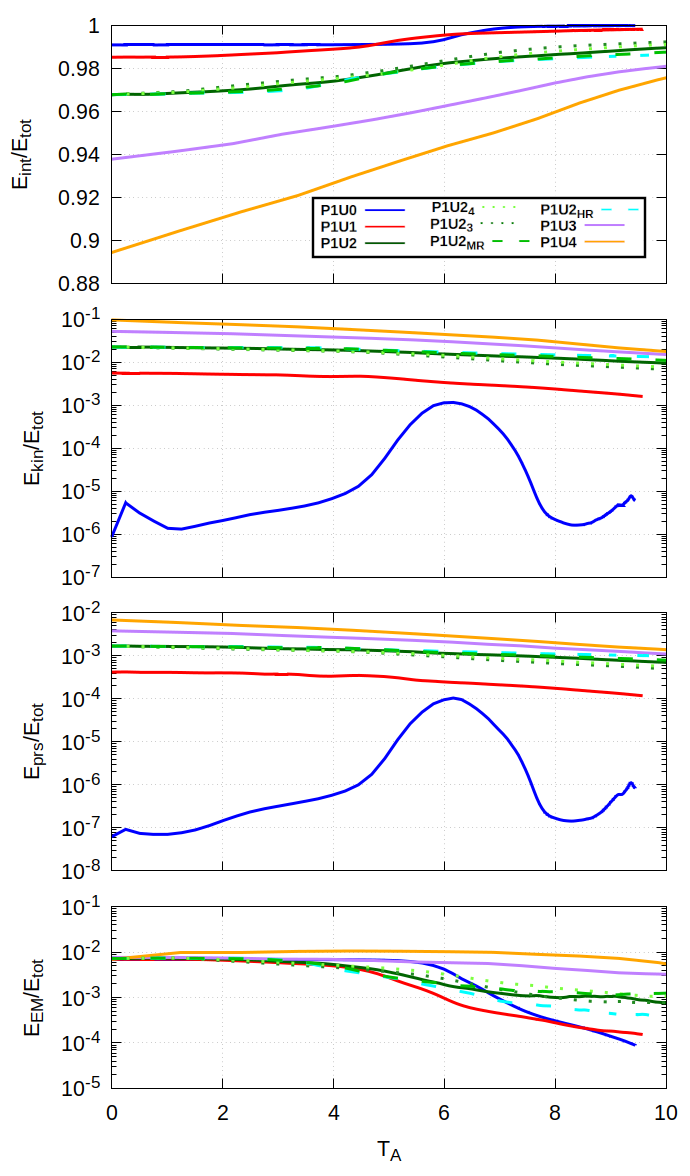}
\caption{Evolution of the internal energy $E_{\rm int}$ (top panel), 
the kinetic $E_{\rm kin}$ energy (second panel), the pressure energy 
$E_{\rm prs}$ (third panel), and the electromagnetic energy $E_{\rm EM}$ 
(bottom panel) normalized by $E_{\rm tot}$ for all simulations of the models listed in 
Table~\ref{tab:velocities}, along with those models in appendices~\ref{sec:atm_study} 
and~\ref{sec:res_study}.}
\label{fig:all_energies}
\end{figure}

Finally, Fig.~\ref{fig:all_energies} displays the evolution of the internal energy, the kinetic energy, the pressure-contribution energy, 
and the electromagnetic energy, all normalized to the total energy defined in Eq.~\eqref{eq:total_energy}, and plotted in units of the 
Alfv\'en crossing time. Over time, the kinetic (except for P1U0), pressure, and electromagnetic components decrease, while most of the total energy is 
gradually converted into internal energy. By the end of the simulations, internal energy is the only component that consistently increases 
across all models except for P1U0, the non-rotating case (see Table~\ref{tab:velocities}), where both the $E_{\rm kin}$ and the $E_{\rm prs}$ 
also increase. This is likely caused by the spontaneous emergence of the differential rotation (see Fig.~\ref{fig:omega_diff}). However, 
their final values remain roughly two orders of magnitude smaller than those in the rotating models, where $E_{\rm kin}$ and $E_{\rm prs}$ 
both decrease. We note that the total energy $E_{\rm tot}$ (not shown) is conserved to within $1\%$ throughout the evolution.
%
%
%
%
\subsection{Rotational timescale effects\label{sec:roteffetc}}
Some instabilities can persist on a timescale of order $\tau_{\mathrm{rot}} = \tau_{\mathrm{A}}^2/T_{\rm r}$ when the rotation 
period $T_{\rm r}$ is shorter than the Alfv\'en time $\tau_{\rm A}$. This was first pointed 
out by~\cite{Pitts} and later discussed in more detail by~\cite{1999A&A...349..189S}. The timescale $\tau_{\rm rot}$ is equivalent 
to the $\tau_{\rm{MHD}}$ of the Flowers-Ruderman mechanism~\citep{Florwers-ruderman}, up to a factor of 2. To account for this timescale in our 
analysis, we define the Alfv\'en rotational crossing time as
\begin{equation}
T_{\mathrm{rot}}=\int_{0}^{t}\frac{1}{\tau_{\mathrm{A}}\sigma_{\mathrm{rot}}}\dd t,
\label{eq:rotalf}
\end{equation}
with $\sigma_{\mathrm{rot}}=\tau_{\mathrm{A}}/T_{\rm r}$ when $\tau_{\mathrm{A}}>T_{\rm r}$, or 
$\sigma_{\mathrm{rot}}=1$ when $\tau_{\mathrm{A}}\leq T_{\rm r}$ \footnote{In this work, we take the 
initial rotation periods in Table~\ref{tab:velocities} for the Alfv\'en rotational crossing time. 
A more refined treatment would be to define an average rotation period at all times. However, we argue that 
the main behavior for $T_{\mathrm{rot}}$ is dominated by $\tau_{\mathrm{A}}$ since, in Eq.~\eqref{eq:rotalf}, 
it is squared when $\tau_{\mathrm{A}}>T_{\rm r}$ (and the changes are more significant).}. We note that of the four rotating models, 
only P1U1 satisfies the condition $\tau_{\mathrm{A}}\leq T_{\rm r}$, and this occurs only at the beginning 
of the simulation before the Alfv\'en time increases.

\begin{figure}
\includegraphics[scale=0.36]{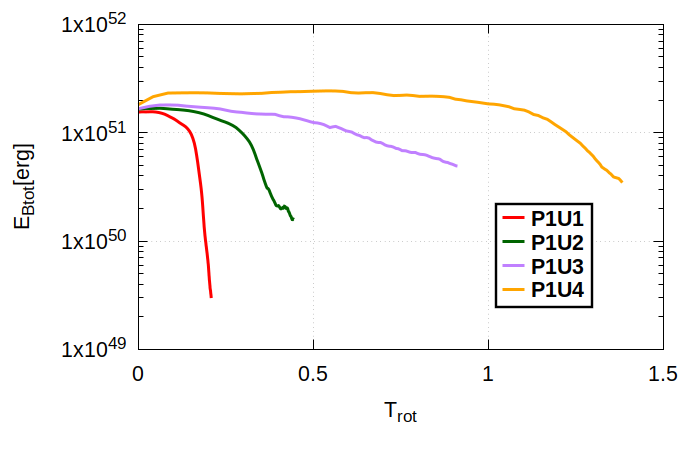}
\caption{Total magnetic energy $E_{B_{\rm tot}}$ vs. the Alfv\'en rotational crossing period $T_{\mathrm{rot}}$.
\label{fig:rotenergy}}
\end{figure}

Fig.~\ref{fig:rotenergy} shows the evolution of magnetic energy versus $T_{\mathrm{rot}}$ for all our rotating models.
Notice that, as our goal is to assess whether rotation delays the onset of instabilities,
we do not shift the time by $T$.
When expressed in units of $T_{\rm{rot}}$, the differences in evolution times between the models become more pronounced.
In particular, P1U4 approaches $T_{\rm{rot}} = 1.5$, while P1U1 does not even reach $T_{\rm{rot}} = 0.25$.
This significant difference arises because the integral in Eq.~\eqref{eq:rotalf} involves $\tau_{\mathrm{A}}^2$, when 
$\tau_{\rm{A}} > T$, unlike Eq.~\eqref{eq:alf}, and also due to the fact that the rotation period $T_{\rm r}$ varies between models.

For P1U1 and P1U2, it appears that the timescale required for instabilities to develop is shorter than $\tau_{\mathrm{rot}}$, 
suggesting that the system becomes unstable on an Alfv\'en timescale instead. In contrast, P1U3 and P1U4 seem to remain stable 
throughout several $\tau_{\mathrm{A}}$, raising the question of whether they can also survive over the longer $\tau_{\rm{rot}}$ 
timescale. The energy decay observed in P1U3 may be attributed to numerical dissipation, while for P1U4, the decay may be 
also numerical before $T_{\rm{rot}}\simeq  1$, after which the steeper drop indicates the onset of an instability 
associated with $\tau_{\rm{rot}}$.

Figure \ref{fig:lines} shows that P1U4 develops a kink instability by $t = 12.5\,T_{\mathrm{A}}$. This suggest that rotation 
delays the onset of the instability but does not suppress it. We note that the resulting energy decline is more gradual than 
in the non-rotating model P1U0 (see Fig.~\ref{fig:casecombo2}). In units of~$\tau_{\mathrm{rot}}$, the evolution of P1U3 terminates 
earlier than that of P1U4, which may explain why we do not yet observe a similarly steep decay. In Fig.~\ref{fig:rotenergy}, 
its energy decrease is more gradual, though it begins to steepen slightly after $T_{\rm{rot}} = 0.5$. Longer simulations than 
those reported here are required to determine whether the system will exhibit a faster decay than P1U4 on this timescale.

%
\subsection{Magnetic field misalignment  
\label{sec:tilting}}
\begin{figure*}
\begin{tabular}{@{}l@{\hspace{0.4mm}}c@{\hspace{0.4mm}}c@{\hspace{0.4mm}}c@{\hspace{0.4mm}}c@{}}
    {} 
    \rotatebox{90}{P1U4} &
        \includegraphics[width=0.242\textwidth]{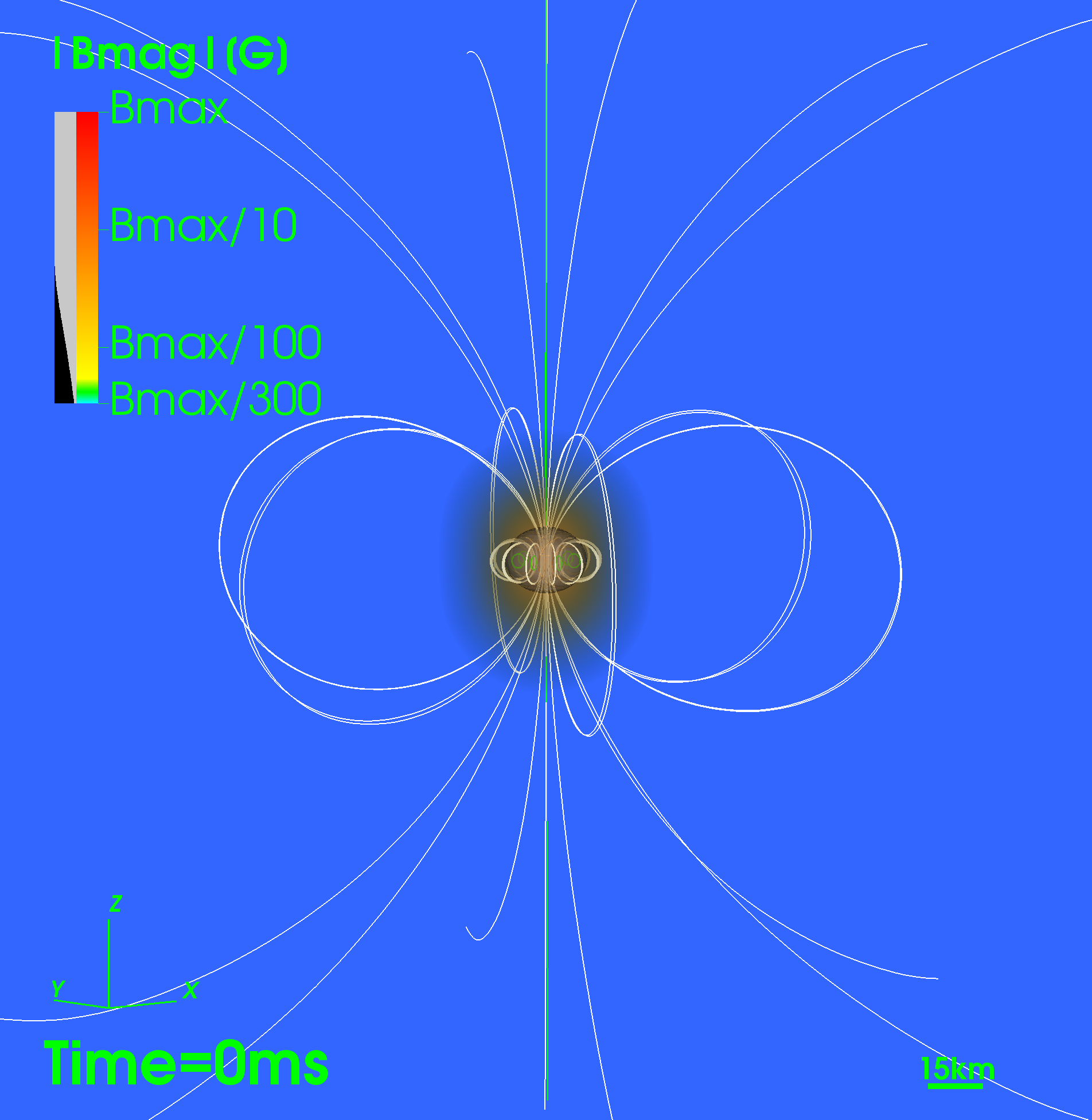} &
        \includegraphics[width=0.242\textwidth]{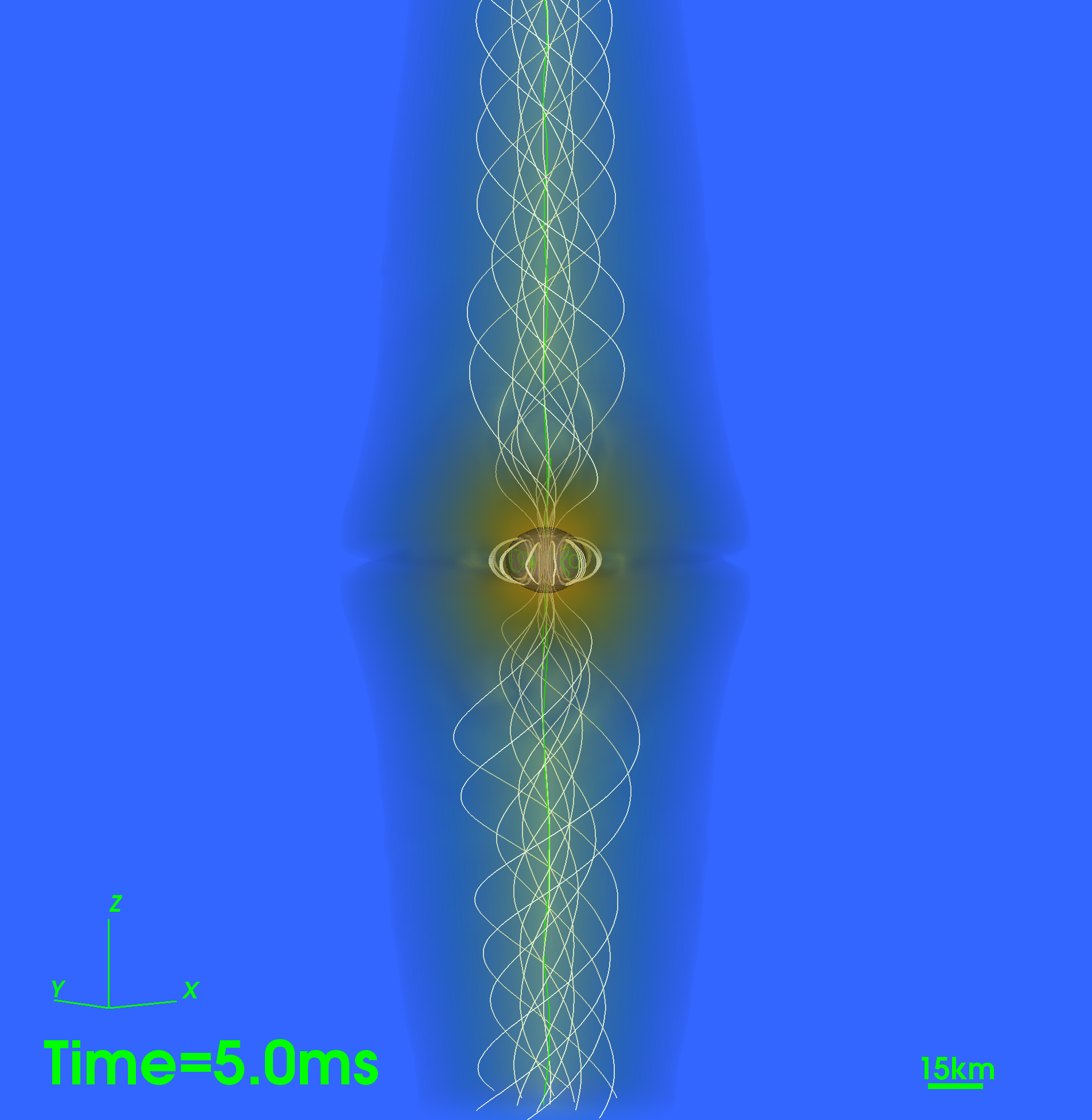} &
        \includegraphics[width=0.242\textwidth]{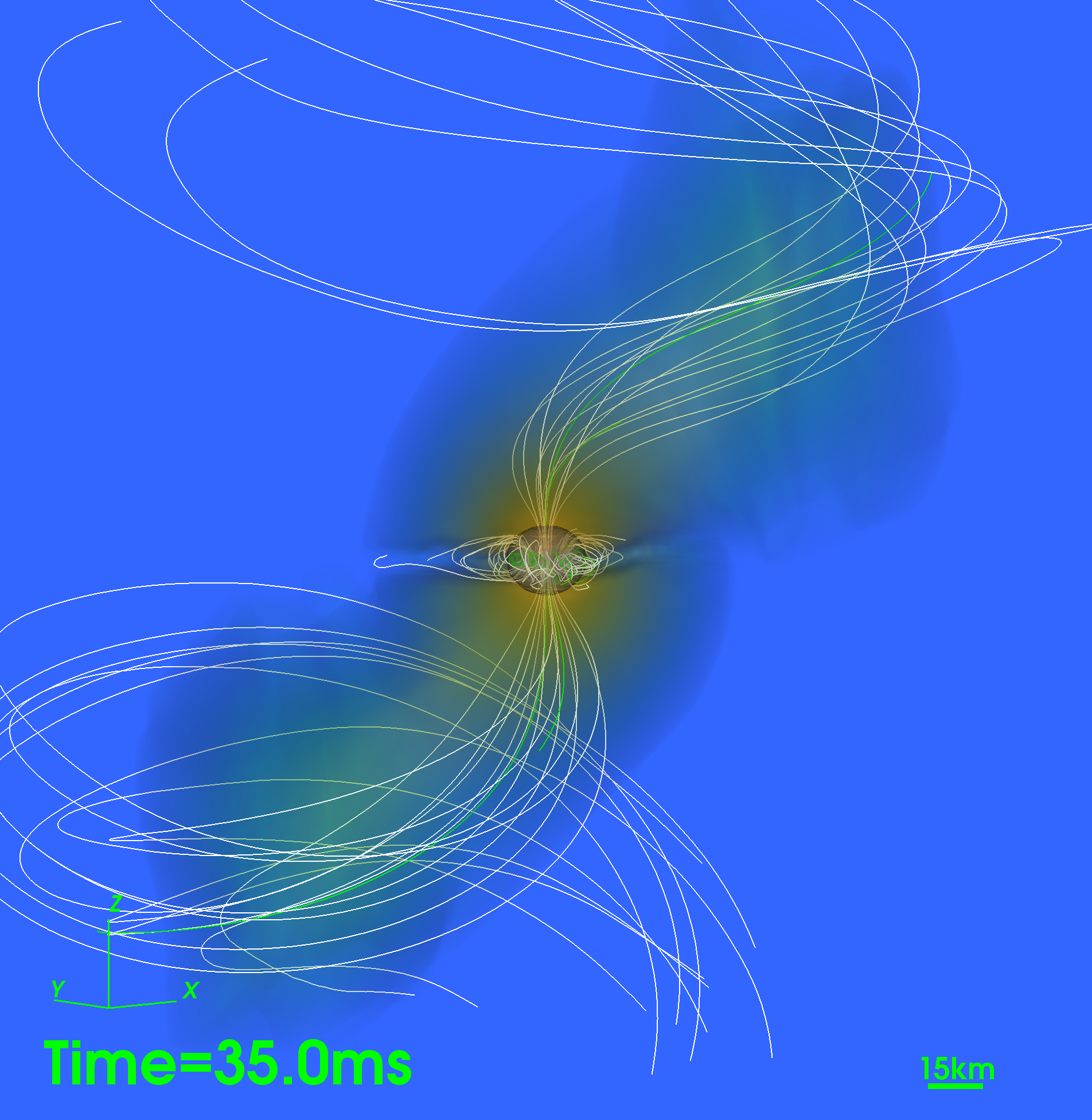} &
        \includegraphics[width=0.242\textwidth]{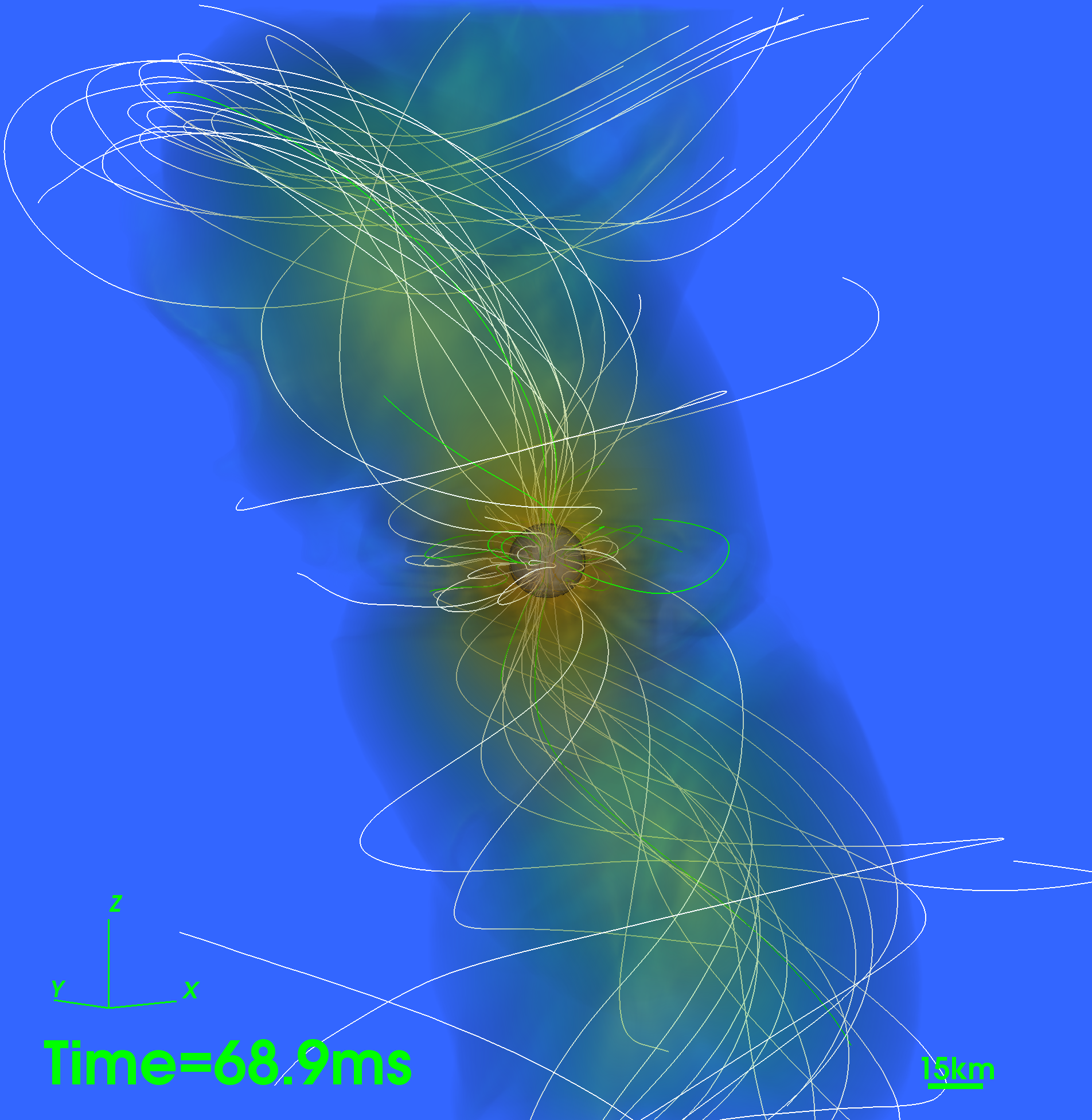} \\
    \end{tabular}
\caption{Strength of the magnetic field for P1U4 at selected times. 
White lines indicate the field lines. Notice that the colorbar limits are rescaled using $B_{\rm max}$ 
at each time to account for changes in field strength. Similar behavior is observed across the other
rotating cases.}
 \label{fig:P1U4_3D_tilt}
\end{figure*}

Although in all of our rotating configurations the NS initial dipole magnetic moment is aligned with 
its angular momentum (see Table~\ref{tab:velocities}), the system naturally evolves toward a configuration in 
which the dipole magnetic moment becomes misaligned with respect to the angular momentum. Fig.~\ref{fig:P1U4_3D_tilt}
displays the strength of the magnetic field at selected times along with the field lines configuration.
%
Initially, the magnetic field strength is axisymmetric along the angular momentum direction and confined to $z\lesssim 2 R_{\rm NS}$
(first panel). As the field is frozen-in, by $t=5~\rm ms$ (or $\sim 1.2T_{\rm{A}}$), it has wound into a helical structure extending 
several NS radii along the angular momentum axis (second panel). However, for $t>5~\rm ms$ (or $\> 7T_{\rm{A}}$), once instabilities 
set in, in particular, the kink instability, the dipole magnetic moment becomes misaligned and appears to precess around the angular 
momentum axis (third and fourth panels). The same behavior is observed in all other rotating cases. A similar effect has been reported,
in the context of binary mergers, during the emergence of magnetically driven jets in~\citep{Ruiz:2021gsv,Ruiz:2021qmm}, where it has
been attributed to the bouncing of magnetically dominated regions as they expand and have overcome the ram pressure of the infalling
material.

To track the precession of the magnetic field, we compute the magnetic field average along the three Cartesian coordinates 
on a sphere $\mathcal{S}$ of coordinate radius $130~\rm km$ and define the tilt angle as
\begin{equation}
  \zeta=\arccos\left(B^z_{\mathrm{avg}}/|B_{\rm avg}|\right)\,.
\label{eq:zeta}
\end{equation}

\begin{figure}[h]
\includegraphics[scale=0.35]{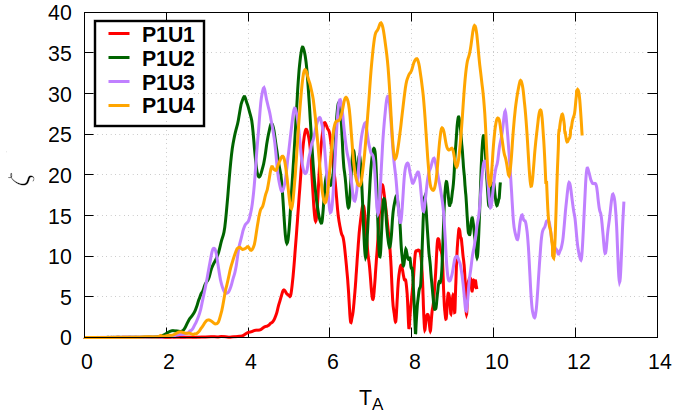}
\includegraphics[scale=0.35]{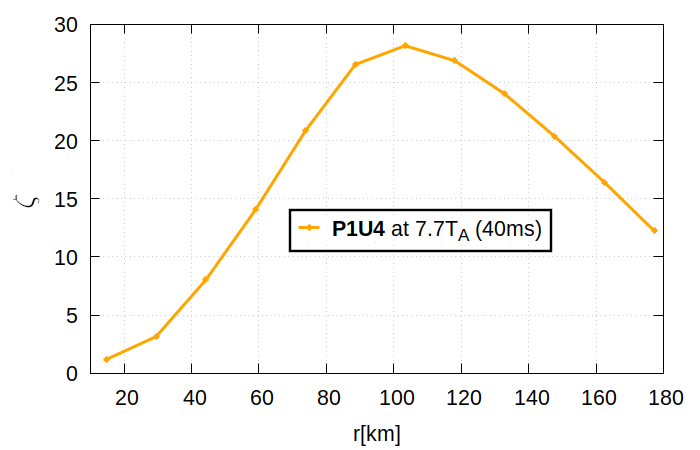}
\caption{Top panel displays the evolution of the tilt angle $\zeta$ for all rotating models
in Table~\ref{tab:velocities} on a sphere of coordinate radius~$130\rm km$. Bottom panel displays
the tilt angle vs. different extraction radii for P1U4 at $7.7T_{\rm{A}}$ (or $40\rm ms$). }
\label{fig:tilting}
\end{figure}
We use Eq.~\eqref{eq:zeta} to estimate the precession of the magnetic field. Top panel in Fig.~\ref{fig:tilting} displays the precession
of the magnetic field strength for all the rotating models. 
As expected, the tilt angle is initially zero, since all models begin with the magnetic dipole moment aligned along the 
direction of the angular momentum. By $t \lesssim 3T_{\rm{A}}$, once some magnetic instabilities have set in, $\zeta$ begins to 
increase, and reaches a maximum peak after the onset of the kink instability. Near to the termination of the simulations, 
$\zeta$ seems to reach a quasi-steady state. The time average of $\zeta$ for each model is shown in Table~\ref{tab:energt_ratio}.
We note that the angle overall increases with angular velocity, 
although a small decrease is observed from PIU2 to PIU3. Longer and higher-resolution simulations are needed to determine 
its asymptotic value and investigate its long-term behavior.
Note that the Eq.~\eqref{eq:zeta} definition is not gauge invariant. The tilt angle $\zeta$ depends on the radius of the 
spherical surface $\mathcal{S}$ over which the average is computed. This is 
illustrated in the lower panel of Fig.~\ref{fig:tilting}, which shows $\zeta$ as a function 
of the coordinate radius for model P1U4 at $\sim 8T_{\rm{A}}$. Close to the star, the tilt angle is relatively small ($\lesssim 3^{\circ}$), 
but it increases with distance, reaching a maximum of approximately $28^{\circ}$, supporting the observation that tilt appears more 
pronounced farther from the star in Fig.~\ref{fig:P1U4_3D_tilt}. Beyond $r = 100\,\rm{km}$, however, $\zeta$ begins to decrease, 
falling to around $12^{\circ}$ near the edge of the simulation domain, where the boundary and the square shape of the grid may influence this value.

%
\section{Discussion and conclusion}
\label{Sec:conc}
The magnetic field stability in NSs remains an open question. Although simple models with purely toroidal or poloidal fields are
unstable~\citep{Braithwaite:2005xi,2013MNRAS.433.2445A}, mixed configurations, especially axisymmetric ones with both components, may be
stable in stars with stable internal layering~\citep{Braithwaite:2008aw,Ciolfi:2009bv}. 
However, there is still no consensus on a universally stable configuration~\citep{2013MNRAS.433.2445A}.

As a step forward in understanding the emergence of magnetic instabilities that disrupt the stellar field configuration, we studied the
impact of stellar rotation using full 3D general relativistic numerical simulations of uniformly rotating NS threaded by strong,
poloidal, pulsar-like magnetic fields. The initial stellar configurations assume perfect conductivity and are stationary and axisymmetric.
We explored a range of angular velocities, from non-rotating stars to those to half the mass-shedding limit.

We found that all configurations spontaneously developed differential rotation, which in turn induces the 
formation of a toroidal magnetic field component that persists until the termination of our simulations. 
By examining the magnetic field structure, we observed that rotation can delay, or even suppress, the emergence 
of some magnetic instabilities. In the non-rotating case (P1U0), we found a clear relation between the magnetic
field energy decay and the growth of the toroidal field, which is also related to the appearance of the kink 
instability. By contrast, in the rotating models this relation becomes less clear, with the decay of magnetic 
field energy appearing later. It is likely that, due to increasing rotation, the instability accompanied by 
the growth of the toroidal magnetic field is delayed. 

Across all models, the total magnetic energy decays within a few Alfv\'en times, although rotation appears to 
slow this decay. We note that in all cases the magnetic energy is still decreasing by the termination of the 
simulations. However, as numerical dissipation is present in all our simulations, it remains difficult to assess 
whether the final magnetic field configurations are stable. 
We found that the faster the rotation, the slower is the decay of the magnetic field energy after the onset of 
the instability, except for the fastest rotating model. This indicates that there may be an optimal period value 
that makes the system retain most of its energy in the Alfv\'en crossing period timescale. This decrease in energy 
for the fastest rotating model is likely related to the instabilities of the timescale $T_{\mathrm{rot}}$ (for this 
timescale, P1U3 simulation is shorter than P1U4). \cite{Pitts,1999A&A...349..189S} reported that rotation 
delays instabilities from the timescale $\tau_{\mathrm{A}}$ to $\tau_{\mathrm{rot}}$ when $T_{\rm r}\ll \tau_{\mathrm{A}}$. 
In our models, this inequality $T_{\rm r} \ll \tau_{\mathrm{A}}$ is poorly achieved by the fastest rotating model
(and even less for the other models) at the beginning of our simulations. Nevertheless, we observe that the instabilities 
for the fastest model are more correlated to this timescale. It is likely that rotation can partially contain the 
instabilities by imposing an axisymmetric shape on the growing toroidal magnetic field, thereby 
preventing axisymmetry from breaking entirely.
In this way, it may suppress the Parker instability and delay the kink instability, both of non-axisymmetric nature. 
Far from the NS, rotation has less effect, and therefore it fails to provide axisymmetry, observed with the magnetic 
field misalignment. On the other hand, as reported in~\citep{Braithwaite_2007}, some angle may be needed between a poloidal 
magnetic field and rotation to interact. In this work, misalignment appears naturally, allowing this interaction, although 
the end states of our simulations are a mixture of poloidal and toroidal magnetic field components so it raises the doubts 
about its importance.

It is worth noting that our simulations do not include a solid-crystalline crust, which could help prevent the dissipation 
of magnetic field energy and may play a crucial role in neutron star stability. We plan to explore this in future work.


\acknowledgements
This work has been supported by the Generalitat Valenciana (grants CIDEGENT/2021/046, Prometeo CIPROM/2022/49 and CIGRIS/2022/126)),
and by the Spanish Agencia Estatal de Investigaci\'on (grants PID2021-125485NB-C21 funded by MCIN/AEI/10.13039/501100011033,
PRE2019-087617, and ERDF A way of making Europe). Further support has been provided by the EU's Horizon 2020 Research and Innovation
(RISE) programme H2020-MSCA-RISE-2017 (FunFiCO-777740) and by the EU Staff Exchange (SE) programme HORIZON-MSCA-2021-SE-01 (NewFunFiCO-101086251).
We acknowledge computational resources and technical support of the Spanish Supercomputing Network through the use of MareNostrum at the Barcelona
Supercomputing Center (AECT-2023-1-0006). A.K.L.Y. acknowledges the computational resources and technical support of the CUHK Central High-Performance
Computing Cluster.  A.K.L.Y. also acknowledges support from the Research Grants Council of Hong Kong (Project No. CUHK 14306419), the Croucher
Innovation Award from the Croucher Foundation Hong Kong, and the Direct Grant for Research from the Research Committee of The Chinese University of
Hong Kong. P.C.-K.C. gratefully acknowledges support from National Science Foundation (NSF) Grant PHY-2020275 (Network for Neutrinos, Nuclear
Astrophysics, and Symmetries (N3AS)).

\appendix \label{appendixas}
%

\section{Impact of a variable magnetosphere on magnetized neutron star evolution}
\label{sec:atm_study}
We note that our density prescription for the MHD evolution of the magnetic field in 
the stellar exterior is only enforced initially. Subsequently, the density is evolved 
everywhere according to the ideal GRMHD equations, with a density floor imposed, as is 
standard in GRMHD simulations~\citep{Ruiz:2018wah,Paschalidis:2014qra}. Empirically, we 
find that stable evolution requires heavier densities for faster-rotating NSs, likely
because rotation induces stronger magnetic gradients.
To explore the impact of the different density prescriptions, ranging from partial to full 
magnetic pressure dominance, we perform simulations of P1U2 with a magnetic-to-gas-pressure
ratio $\beta^{-1}=10^{-3}, 10^{-4}, 10^{-6}$, denoted as $\mathrm{P1U2}_3$, $\mathrm{P1U2}_4$ 
and P1U2, respectively. 

\begin{figure}
\centering{
    \includegraphics[width=\textwidth]{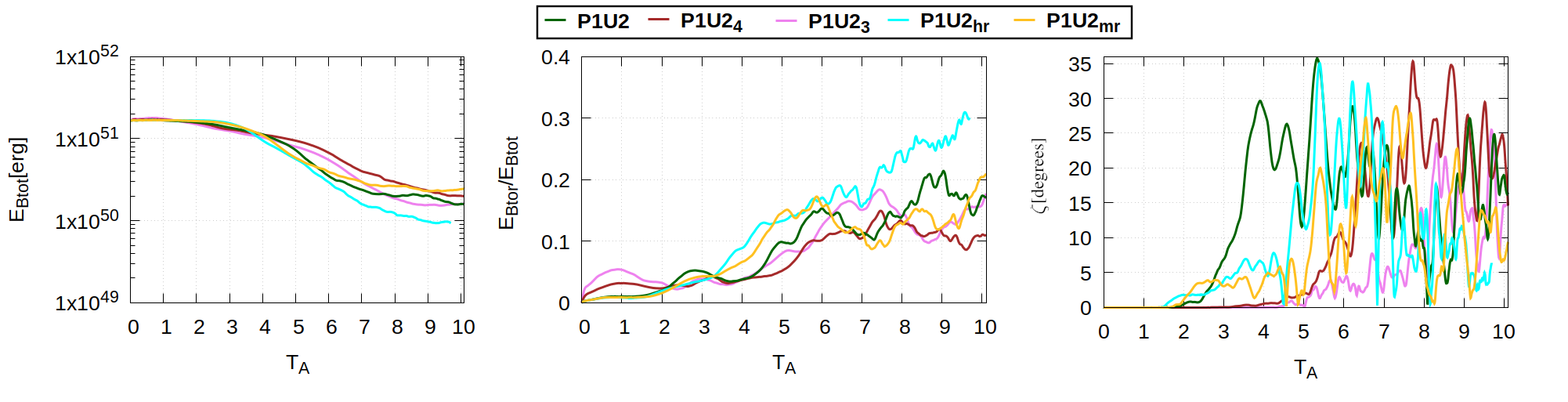}
    \caption{Evolution of the total magnetic energy (left), toroidal magnetic energy (middle), and 
    the angle $\zeta$ between the magnetic field and the rotation axis (right) for P1U2, using three different 
    atmosphere prescriptions, ranging from partial to full magnetic pressure dominance (Appendix~\ref{sec:atm_study}),
    and two higher resolutions (Appendix~\ref{sec:res_study}). 
 \label{fig:atmo_comparison}}}
\end{figure}

The left panel in Fig.~\ref{fig:all_energies} displays the evolution of different energies (see 
Eqs.~\eqref{eq:E_int}-\eqref{eq:E_EM}) for the above cases. We note that, regardless of the initial
density prescription, all energy components evolve following the same trend with no significant 
changes between them. Fig.~\ref{fig:atmo_comparison} shows the evolution of the total magnetic 
energy $E_{B_{\rm tot}}$. We observe that changes induced by the different atmosphere prescriptions
are $\lesssim 3\%$ (see Table~\ref{tab:energt_ratio2}), compared to those with different angular 
velocities, where changes range from $\sim 6\%$ to $\sim 58\%$ (see Table~\ref{tab:energt_ratio}).

As noted earlier, the stars spontaneously develop differential rotation, which leads to the emergence 
of a strong toroidal magnetic field component that persists until the end of our simulations. As shown 
in the middle panel of Fig.~\ref{fig:all_energies}, different atmospheres affect the magnitude of the 
toroidal magnetic field component. Near the end of our simulations, we observe variations of about 
$\sim 8\%$ in the toroidal field strength across the evolved models (see Table~\ref{tab:energt_ratio2}). 
However, by $t\sim 10T_{\rm A}$ the system has not yet reached a steady state, and longer simulations 
are therefore required to fully assess the impact of the atmosphere on the growth of the toroidal 
magnetic field component.

Fig.~\ref{fig:atmo_comparison_diff} shows the angular velocity profiles for these three cases. During the 
early evolution, we observe that models with a heavier atmosphere exhibit sharper differential rotation in 
the inner region of the star. However, after $\sim 10,\rm ms$, the stars roughly recover a uniform
rotation of $\sim 1.25,\rm kHz$ for P1U2, and slightly lower values ($\sim 1.1,\rm kHz$) for the other cases.

Finally, it is worth noting that the tilt angle, shown in Fig.~\ref{fig:atmo_comparison}, is roughly the 
same for cases with $\beta=10^{-4}$ and $10^{-6}$, but changes significantly for the $\beta=10^{-3}$ case 
(see Table~\ref{tab:energt_ratio2}). This is likely due to the larger ram pressure that the field lines must 
overcome. Longer numerical simulations are needed to probe the impact of heavier atmospheres in the tilt angle.
%
%
%
\section{Influence of resolution on magnetic evolution}
\label{sec:res_study}
To assess the dependence of our results on the numerical resolution, we consider two additional P1U2 cases using 
higher resolutions: i)~$\mathrm{P1U2_{MR}}$, where we employ a resolution of $N_x \times N_y \times N_z = 96^3$;
and ii)~$\mathrm{P1U2_{HR}}$, with $N_x \times N_y \times N_z = 128^3$. Both simulations use five AMR levels, 
differing in size and resolution by a factor of two, i.e. we are employing one additional refinement level 
compared to our canonical cases. The finest grid size at the center of the star is $\Delta x = \Delta y = \Delta z
\approx 230~\rm m$ for $\mathrm{P1U2_{MR}}$ and $\Delta x = \Delta y = \Delta z \approx 173~\rm m$ for 
$\mathrm{P1U2_{HR}}$. All other aspects of the two simulations are identical to those of the canonical P1U2 case.
In particular, we set $\beta^{-1}=10^{-6}$ in these two cases.

Fig.~\ref{fig:all_energies} shows the evolution of different energies (see Eqs.~\eqref{eq:E_int}–\eqref{eq:E_EM}). 
In general, we find that numerical resolution has only a minor impact on all energies, with variations smaller than $2\%$, 
except $E_{\rm EM}$. Although the overall trend in $E_{\rm EM}$ is consistent across all cases, it exhibits a more 
pronounced decay in $\mathrm{P1U2_{HR}}$ than in P1U2. By contrast, the electromagnetic energy in $\mathrm{P1U2_{MR}}$ 
and P1U2 remains roughly the same (see bottom panel of Fig.~\ref{fig:all_energies}). We also note that since this energy 
is roughly an order of magnitude smaller than the other energies, its behavior only marginally affects the internal energy.

\begin{figure}
\centering{
    \includegraphics[width=\textwidth]{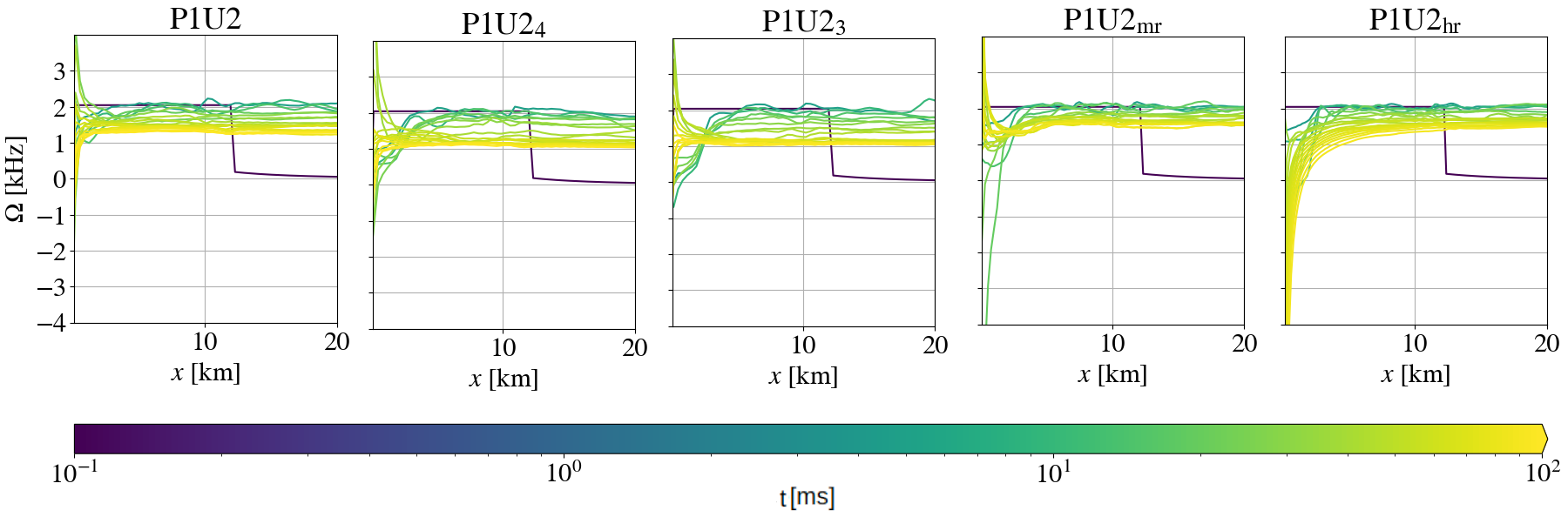}
    \caption{Angular velocity profiles along the coordinate x-axis for P1U2 using three different 
    atmosphere prescriptions, ranging from partial to full magnetic pressure dominance (Appendix~\ref{sec:atm_study}),
    and two higher resolutions (Appendix~\ref{sec:res_study}).
 \label{fig:atmo_comparison_diff}}}
\end{figure}

The left panel of Fig.~\ref{fig:atmo_comparison} shows the evolution of the total magnetic energy $E_{B_{\rm tot}}$.
During the first $t\sim 4T_{\rm A}$, numerical resolution has a minor impact on its evolution. However, after the onset of 
magnetic instabilities, a stronger toroidal magnetic field component develops in the highest-resolution case (see middle
panel in Fig.~\ref{fig:atmo_comparison}). As shown in Fig.~\ref{fig:atmo_comparison_diff}, differential rotation is also 
more pronounced in this case, leading to a more rapid decay of the total magnetic energy than in the other cases 
(see Table~\ref{tab:energt_ratio2}). In particular, the energy decays by a factor of $\sim 1.8$ in $\mathrm{P1U2_{HR}}$ 
compared to P1U2. By contrast, in $\mathrm{P1U2_{MR}}$ the total magnetic energy near the end of the simulations is about 
$1.3$ times larger than in P1U2. Similar behavior has been reported in~\citep{Sur_2022}.

Finally, the tilt angle $\zeta$, shown in the right panel of Fig.~\ref{fig:atmo_comparison}, is slightly smaller in the 
high-resolution cases than in the canonical P1U2 (see Table~\ref{tab:energt_ratio2}). This suggests that numerical viscosity 
may play a significant role in the precession of the magnetic field lines, indicating that higher-resolution simulations are 
required to assess its impact on the tilt angle.
%
%
%
%
\begin{table} 
    \centering
     \caption{Comparison of the evolution of P1U2 evolved using different
     atmosphere prescription and higher resolutions. We list the fraction of 
     the initial total magnetic energy $E_{B_{\rm tot}}$ remaining around 
     the termination of our simulations ($\sim 9.5T_{\rm A}$), the toroidal-to-total magnetic 
     energy ratio at that point, and the magnetic field tilt angle with respect 
     to the rotation axis for all NS models. 
    \label{tab:energt_ratio2}}
    \scalebox{0.9}{%
    \begin{tabular}{c|cccccccc}
\hline
\hline
        MODEL &  $E_{B_{\rm tot}}/E_{B_{\rm tot}0}$ &  $E_{{B_{\rm tor}}}/{E_{B_{\rm tot}}}$ & $\zeta_{\mathrm{avg}}$[º] \\
\hline
        P1U2                 &   $10.4\%$ &  $16.9\%$ & $14.6$  \\
        $\rm{P1U2}_4$        &   $12.6\%$ &  $9.3\%$  & $14.6$  \\
        $\rm{P1U2}_3$        &   $9.3\%$  &  $14.2\%$ & $7.8$  \\
        $\rm{P1U2}_{\rm MR}$ &   $13.6\%$ &  $11.1\%$ & $10.4$  \\
        $\rm{P1U2}_{\rm HR}$ &   $5.8\%$  &  $29.6\%$ & $10.1$  \\
\hline
\hline
    \end{tabular}}
\vspace{7pt}
\end{table}

\bibliographystyle{hapj}
\bibliography{ref}

\end{document}

%% file: kigou.tex
\newcommand{\beq}{\begin{equation}} 
\newcommand{\eeq}{\end{equation}} 
\newcommand{\beqn}{\begin{eqnarray}} 
\newcommand{\eeqn}{\end{eqnarray}}

\newcommand{\zD}{{\raise1.0ex\hbox{${}^{\ \circ}$}}\!\!\!\!\!D}
\newcommand{\alone}{{\raise0.5ex\hbox{${}^{\ 1}$}}\!\!\!\!\alpha}

\newcommand{\nalam}{\mathrel{\raise0.9ex\hbox{$^\lambda$}\mkern-14mu
\lower0.0ex\hbox{$\nabla$}}}

\newcommand{\zeroD}{{\raise1.0ex\hbox{${}^{\ \circ}$}}\!\!\!\!\!D}

\newcommand{\zLap}{{\raise1.0ex\hbox{${}^{\ \circ}$}}\!\!\!\!\Delta}
\newcommand{\zna}{{\raise1.0ex\hbox{${}^{\ \circ}$}}\!\!\!\!\!\nabla}
\newcommand{\zS}{{\raise1.0ex\hbox{${}^{\ \circ}$}}\!\!\!\!\!S}

